\documentclass[12pt,english]{article}
\usepackage[T1]{fontenc}
\usepackage[utf8]{inputenc}
\usepackage{geometry}
\usepackage{babel}
\usepackage{array}
\usepackage{verbatim}
\usepackage{textcomp}
\usepackage{amsmath}
\usepackage{graphicx}
\usepackage[unicode=true,pdfusetitle,
 bookmarks=true,bookmarksnumbered=false,bookmarksopen=false,
 breaklinks=false,pdfborder={0 0 0},pdfborderstyle={},backref=false,colorlinks=false]
 {hyperref}

\makeatletter

\let\SF@@footnote\footnote
\def\footnote{\ifx\protect\@typeset@protect
    \expandafter\SF@@footnote
  \else
    \expandafter\SF@gobble@opt
  \fi
}
\expandafter\def\csname SF@gobble@opt \endcsname{\@ifnextchar[
  \SF@gobble@twobracket
  \@gobble
}
\edef\SF@gobble@opt{\noexpand\protect
  \expandafter\noexpand\csname SF@gobble@opt \endcsname}
\def\SF@gobble@twobracket[#1]#2{}
\providecommand{\tabularnewline}{\\}

\numberwithin{equation}{section}
\numberwithin{figure}{section}
\numberwithin{table}{section}
\newcommand{\lyxaddress}[1]{
	\par {\raggedright #1
	\vspace{1.4em}
	\noindent\par}
}

\renewcommand\[{\begin{equation}}
\renewcommand\]{\end{equation}}
\@addtoreset{equation}{section}

\@ifundefined{showcaptionsetup}{}{%
 \PassOptionsToPackage{caption=false}{subfig}}
\usepackage{subfig}
\makeatother

\begin{document}
\title{Infrared Thermography in the Tokamak à Configuration Variable}
\author{Martim Zurita\thanks{martim.zurita@epfl.ch}, H. Reimerdes, C. Colandrea,
H. Elaian, M. Pedrini,\\
Y. Andrebe, F. Crisinel, S. Koncewiez, J.-D. Landis, D. Mykytchuk,\\
U. Sheikh, and the TCV team\textbf{}\thanks{Author list at the end of the paper.}}
\maketitle

\lyxaddress{\begin{center}
École Polytechnique Fédérale de Lausanne (EPFL), Swiss Plasma Center
(SPC), \\
CH-1015 Lausanne, Switzerland
\par\end{center}}
\begin{abstract}
In the Tokamak à Configuration Variable (TCV), infrared thermography
(IR) is currently composed of the horizontal, vertical, and tangential
infrared systems (HIR, VIR, TIR), which all use Equus 81k M cameras.
The IR diagnostics obtain the surface temperature of TCV\textquoteright s
graphite tiles for post-discharge analysis. Target heat flux profiles
are inferred from the tile temperature with the THEODOR (\textbf{Th}ermal
\textbf{E}nergy \textbf{O}nto \textbf{D}ivert\textbf{or}) code. Fast
transient analysis is possible in reduced frame mode, with acquisition
frequencies above 10kHz. The main views are the lower inner wall for
HIR, the floor for VIR, and the lower outer wall for TIR. The HIR
camera can also be moved to view the midplane inner wall, while TIR
can be moved to see the midplane inner wall and the upper outer wall,
mainly to measure synchrotron radiation and heat deposition due to
runaway electrons. Recent developments in TCV\textquoteright s IR
systems include (i) tile diffusivity and conductivity measurements
to assure the precision of heat flux estimates; (ii) the addition
of one new VIR heated valley tile and two rooftop TIR tiles, for measurements
of fast heat flux transients; (iii) the implementation of long-pass
wavelength filter of 4095 nm, to diminish the measurement of plasma
parasitic infrared light, mainly from deuterium $5\rightarrow4$ emission
at 4051 nm. Despite these developments, the main sources of uncertainty
for IR in TCV are still parasitic infrared light and the determination
of the surface layer heat transmission factor, both of which mainly
affect the VIR system.

\pagebreak
\end{abstract}
\tableofcontents{}

\pagebreak

\section{Introduction}

Infrared thermography (IR) is an imaging technique that translates
the infrared radiation emitted by an object into its surface temperature.
This is done by employing the total photon radiance from blackbody
theory, $R_{BB}$ \cite{Gaussorgues IR book}:

\begin{equation}
R_{BB}(T)=\int_{\lambda_{1}}^{\lambda_{2}}\frac{2c}{\lambda^{4}}\frac{d\lambda}{\exp\left(\frac{hc}{\lambda k_{B}T}\right)-1}\quad\left[\frac{\text{1}}{\text{sr}\;\text{s}\;\text{m}^{2}}\right]\label{eq: R_BB(T)}
\end{equation}
where $c=2.99792\times10^{8}\;\text{m/s}$ is the light speed, $h=6.62607\times10^{-34}\;\text{J s}$
is Planck's constant, $k_{B}=1.38065\times10^{-23}\;\text{J/K}$ is
Boltzmann's constant, $T$ is the object surface temperature, and
$\lambda_{1}$ and $\lambda_{2}$ are the minimum and maximum wavelength
measured by the camera detector. $R_{BB}(T)$ measures the number
of photons per solid angle, time and area emitted by the object. Inverting
$R_{BB}(T)$ gives the surface temperature as a function of the infrared
radiation. In tokamaks and stellarators, infrared thermography is
used to evaluate the temperature of the device walls in contact with
the plasma. By employing heat diffusion equations, the heat flux due
to the plasma can then be estimated. 

This paper details the IR systems of the \emph{Tokamak à Configuration
Variable} (TCV) \cite{Duval 2024 TCV overview}. Sec. \ref{sec: IR in TCV}
describes TCV's infrared cameras; Sec. \ref{sec:Calibration-to-temperature},
the calibration to temperature with thermocouples; Sec. \ref{sec:Heat-flux-estimation},
the target heat flux estimation with the THEODOR code; Sec. \ref{sec:Windows},
the windows used for IR in TCV; Sec. \ref{sec:Tiles}, the tiles ---
with emphasis for their thermal properties and new tiles developed
for fast transient analysis; Sec. \ref{sec:IR-plasma-radiation} discusses
the infrared parasitic plasma radiation measured by the cameras; Sec.
\ref{sec:VIR-vs-TIR} shows a comparison between the VIR and TIR heat
flux profiles and Sec. \ref{sec:Summary} summarizes the results.

\section{\label{sec: IR in TCV}TCV\textquoteright s infrared cameras }

\subsection{General information }

All infrared cameras in TCV are currently IRCAM Equus 81k M (specifications
in Table \ref{tab:Gen-Info}), adapted by the manufacturer to withstand
\textquoteleft high\textquoteright{} magnetic fields (0.1-0.2 T).
The detector is a focal plane array (FPA) made of mercury cadmium
telluride (MCT). A Stirling cooler refrigerates the detector, minimizing
its own infrared emission and reducing noise due to thermally excited
electron holes. The sensor is a complementary metal oxide semiconductor
(CMOS), with each pixel having its own amplifier. The length of each
pixel (i.e. the pixel pitch) is of $30\;\mu\text{m}$. The integration
mode of the data is integrate-then-read. 

The cameras\textquoteright{} spectral response goes from $3.7\;\mu\text{m}$
to $4.8\;\mu\text{m}$, falling in the mid-wavelength infrared (MWIR)
range. Nevertheless, as of TCV\textquoteright s discharge 79993, all
three cameras are filtered with a $4095\;\text{nm}$ long-pass wavelength
filter to decrease the amount of infrared light measured from the
plasma. 

The maximum and minimum frame sizes are respectively 320x256 (approximately
81 thousand pixels, hence the camera name Equus 81k M) and 64x2 pixels.
Decreasing the frame size increases the acquisition frequency, as
described in Eq. (\ref{eq: dt frame min 1}). The spatial resolution
ranges from about 2.0mm/pixel (VIR and TIR) to 1.3mm/pixel (HIR). 

The minimum integration time (i.e. the interval the camera pixels
are exposed to light) is $\Delta t_{int}=1\;\text{\ensuremath{\mu}s}$,
which can be used for cases with very high tile temperatures (e.g.
HIR of discharge 89000, not shown). Nevertheless, the standard value
is $\Delta t_{int}=1\;\text{ms}$ for L-mode, ohmically heated plasmas. 

TCV's infrared cameras are controlled by a C++ program with MDS events
(Sec. 3.1 of \cite{Maurizio 2020 PhD thesis}). As of 2025, the quality
of the data is monitored shot by shot in TCV's DdJ panel \cite{Molina 2026 DdJ}.

\begin{table}
\caption{\label{tab:Gen-Info}General information about TCV\textquoteright s
IR cameras Equus 81k M by IRCAM.}

\centering{}%
\begin{tabular}{|c|c|}
\hline 
Detector & MCT FPA\tabularnewline
\hline 
Sensor & CMOS\tabularnewline
\hline 
Integration mode & Integrate-then-read\tabularnewline
\hline 
Wavelength sensitivity & $3.7\;\mu\text{m}$ (or $4.1\;\mu\text{m}$ filtered) to $4.8\;\mu\text{m}$\tabularnewline
\hline 
Max window size & 320x256 pixels\tabularnewline
\hline 
Min window size & 64x2 pixels\tabularnewline
\hline 
Pixel pitch & $30\;\mu\text{m}$\tabularnewline
\hline 
Spatial resolution & 1.3mm/pixel (HIR) and 2mm/pixel (VIR \& TIR)\tabularnewline
\hline 
Integration time & $1\;\mu\text{s}$ to $10\;\text{ms}$\tabularnewline
\hline 
Acquisition frequency & 200 Hz (standard) to 20kHz (max)\tabularnewline
\hline 
\end{tabular}
\end{table}

\subsection{Interval between frames}

As described in the cameras user manual and in Sec. 3.1.2 of \cite{Maurizio 2020 PhD thesis},
the minimum interval between frames of TCV's  81k M IR cameras is

\begin{align}
\Delta t_{frame}^{(min)} & =\Delta t_{int}+\Delta t_{read}+\Delta t_{reset}\label{eq: dt frame min 1}\\
 & =\Delta t_{int}+N_{px}/(4f_{clk})+20/f_{clk}\nonumber 
\end{align}
where $\Delta t_{int}$ is the integration time, i.e. the interval
in which the camera receives light for each frame. $\Delta t_{read}=N_{px}/(4_{clk})=N_{px}/27$
is the readout time of the data, $N_{px}$ is the number of pixels
used by the camera (max $320\times260=81920$, min $64\times2=128$,
as in Table \ref{tab:Gen-Info}) and $f_{clk}=6.75\;\text{MHz}$ is
the camera mother-clock frequency. $\Delta t_{reset}=20/f_{clk}=20/(6.75\;\text{MHz})=3\;\mu\text{s}$
is estimated as the time to reset the pixels for the next measurements.
It is also possible to change $f_{clk}$ to $10\;\text{MHz}$, decreasing
$\Delta t_{read}$ by 32\%. However, this leads to pixels interfering
to each other (pixel cross-talk), which can distort the data.

Substituting $f_{clk}=6.75\;\text{MHz}$ in Eq. (\ref{eq: dt frame min 1})
gives the general formula for the minimum interval between frames

\begin{equation}
\Delta t_{frame}^{(min)}=\Delta t_{int}+\frac{1}{27}N_{px}\;\mu\text{s}+3\;\mu\text{s}\label{eq: dt frame min 2}
\end{equation}
In full frame mode, $N_{px}=320\times260=81920$, yielding $\Delta t_{read}=3.0\;\text{ms}$
and hence 
\[
\Delta t_{full\;frame}^{(min)}=\Delta t_{int}+3.0\;\text{ms}
\]
Since in general $\Delta t_{int}<3\;\text{ms}$, the full frame interval
is dominated by the readout time and the maximum acquisition frequency
is then obtained for $\Delta t_{int}\rightarrow0$, $f_{acq}=1/\Delta t_{full\;frame}^{(min)}=333\;\text{Hz}$.

In practice, to assure reliable measurements, the used interval between
frames is increased by a factor of 10\% and an adjustment $\Delta t_{adjust}$
is added,

\begin{align}
\Delta t_{frame} & =(\Delta t_{int}+\Delta t_{read}+\Delta t_{reset}+\Delta t_{adjust})\cdot1.1\label{eq: dt frame final}\\
 & =(\Delta t_{int}+N_{px}/27\;\mu\text{m}+3\;\mu\text{s}+15\;\mu\text{s})\cdot1.1\nonumber 
\end{align}
$\Delta t_{adjust}=15\;\mu\text{s}$ is important for low number of
pixels ($N_{px}<2400$), specially at extremely low integration times
($1\sim5\;\mu\text{s}$) to assure the identification of the frame
instant. Decreasing the readout time too much can also lead to frames
not being recorded and thus a minimum of $N_{px}=64\times24=1536$
is recommended. 

\subsection{HIR: the horizontal infrared system}

\begin{figure}
\begin{centering}
\subfloat[]{\begin{centering}
\includegraphics[scale=0.6]{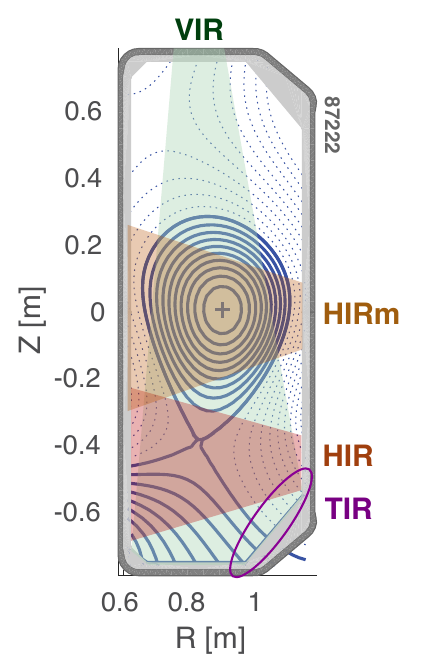}
\par\end{centering}
}\subfloat[]{\begin{centering}
\includegraphics[scale=0.5]{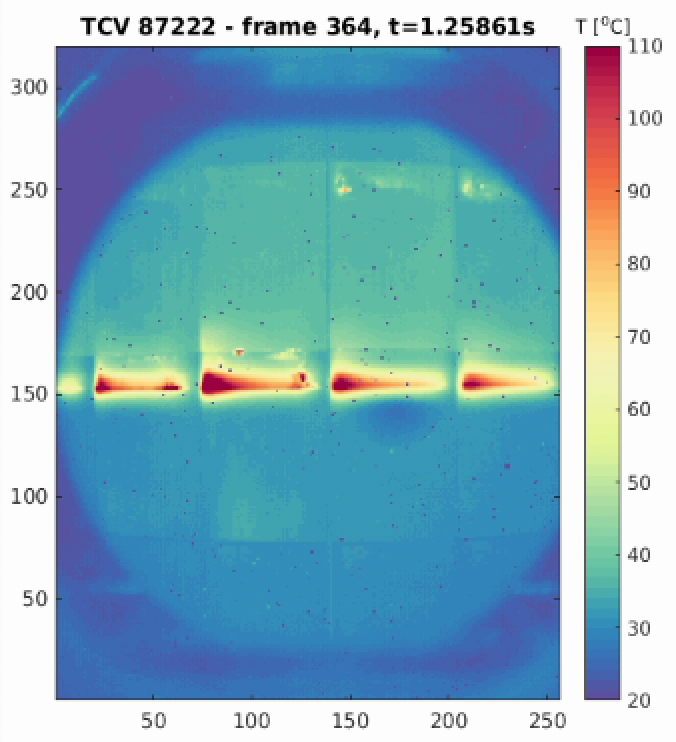}
\par\end{centering}
}\subfloat[]{\begin{centering}
\includegraphics[scale=0.5]{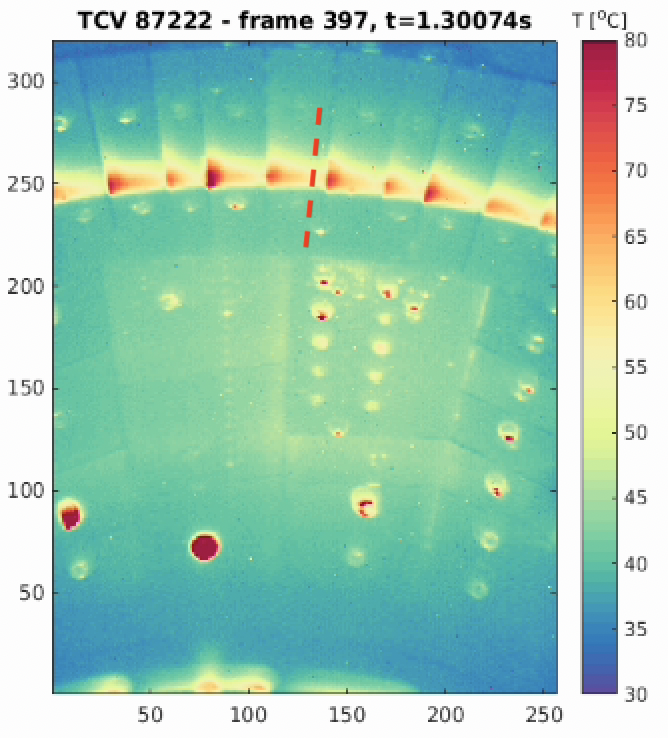}
\par\end{centering}
}
\par\end{centering}
\caption{\label{fig:HIR =000026 VIR view}(a) Magnetic reconstruction of a
power exhaust reference shot (87222), repeated from time to time to
assess the divertor conditions, with all IR camera views except the
TIR upper port. (b) HIR view, with inner strike point at the inner
wall. Differences in temperature in the strike point are due to the
inner wall tiles not being toroidally symmetric \cite{Pitts1999 TCV tiles}.
(c) VIR view with outer strike point (OSP) at the tilted tiles and
line of interest (in red) used for the heat flux analysis in Fig.
\ref{fig: heat flux example}. Red and yellow circles are screws.
Difference in temperature in the OSP are due to leading edges and
absence of toroidal symmetry.}

\begin{centering}
\subfloat[]{\begin{centering}
\includegraphics[scale=0.5]{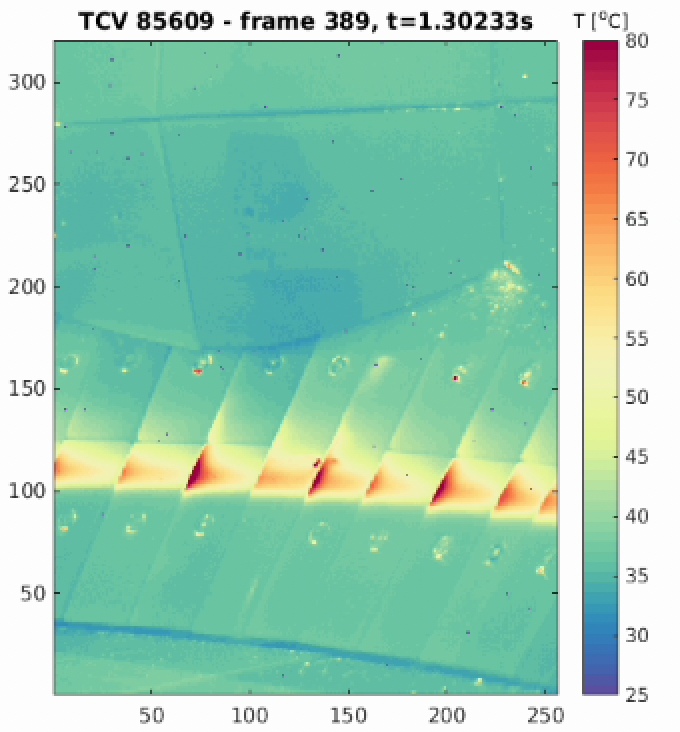}
\par\end{centering}
}\subfloat[]{\begin{centering}
\includegraphics[scale=0.5]{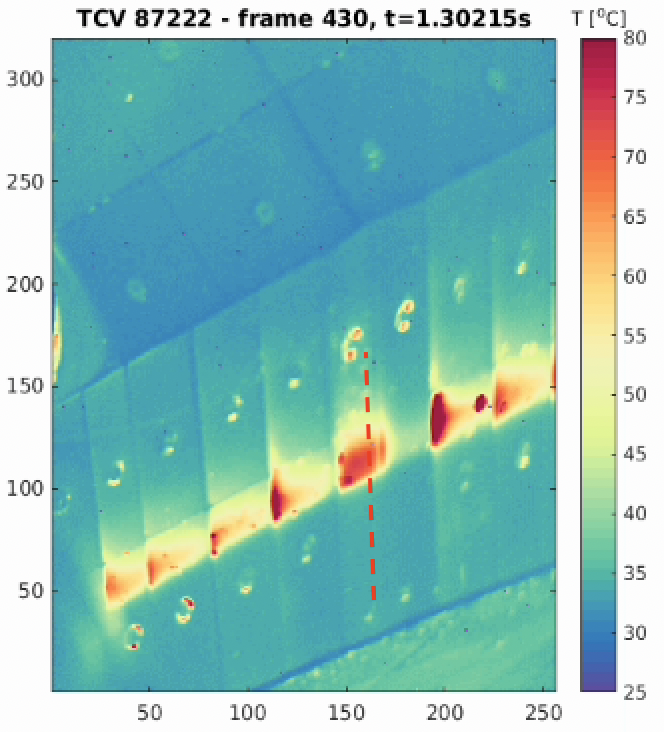}
\par\end{centering}
}
\par\end{centering}
\caption{\label{fig: TIR view}TIR main view before (a) and after (b) being
rotated by 28º, during two identically programmed PEX reference shots
with OSP at the tilted tiles. The dashed line indicates the line of
interest used to estimate the target heat flux in Fig. \ref{fig: heat flux example}.
The discharge 87222 also already had the new rooftop tile installed.
Higher temperature at the edges happens due to leading edges and the
fact that the tilted tiles are not toroidally symmetric.}
\end{figure}

The HIR standard view is the lower port of TCV\textquoteright s graphite
tiles, where the inner strike point (SP) of most of TCV\textquoteright s
plasmas is located (Fig. \ref{fig:HIR =000026 VIR view}). HIR is
also routinely moved to the midplane (HIRm), to measure the inner
SP of long-legged configurations, synchrotron radiation from runaway
electrons, and RE heat deposition. The history of HIR is described
in Table \ref{tab:HIR history}.

\begin{table}
\caption{\label{tab:HIR history}HIR history. }

\begin{centering}
\begin{tabular}{|>{\centering}p{3cm}|>{\centering}p{12cm}|}
\hline 
Date & Event\tabularnewline
\hline 
\hline 
15/7-7/9/09 & HIR camera borrowed from MAST\tabularnewline
\hline 
2012 & Decoupled HIR support from tokamak to avoid vibration\tabularnewline
\hline 
2013 & Mid-wavelength camera used for HIR (1st shot: 48508)\tabularnewline
\hline 
2015 & IRCAM Equus1 81k M camera acquired and now uses a new and a 25mm lens;
re-allocated from TCV sector 4 to 7; can be placed on the midplane
port or a lower port\tabularnewline
\hline 
Dec. 2015 & Purchase of a 12.5mm lens from IRCAM\tabularnewline
\hline 
Jun. 2022 & Midplane sapphire window replaced by germanium\tabularnewline
\hline 
23 Nov. 2022 & Midplane germanium window cracks\tabularnewline
\hline 
Jan. 2023 & New midplane sapphire window installed\tabularnewline
\hline 
Jul. 2025 & 6mm ZnSl lower port window with AR coating cracked on installation\tabularnewline
\hline 
25 Sep. 2025 & New lower port sapphire window installed (as of discharge 87965)\tabularnewline
\hline 
\end{tabular}
\par\end{centering}
\textcompwordmark{}

\textcompwordmark{}

\caption{\label{tab:VIR history}VIR history. 32951 is not the 1st discharge,
but a discharge of interest.}

\centering{}%
\begin{tabular}{|c|>{\centering}p{11cm}|c|}
\hline 
Date & Event & 1st discharge\tabularnewline
\hline 
\hline 
Oct. 2006 & Thermosensorik camera installed & 32951{*}\tabularnewline
\hline 
Mar. 2008 & Optics v2, with better image quality and wider view & 34756\tabularnewline
\hline 
Mar. 2017 & IRCAM Equus2 81k M replaced Thermosensorik & \tabularnewline
\hline 
09/10/20 & VIR shutter broke & \tabularnewline
\hline 
28/02/21 & Ge window with AR coating replaced old ZnSe window & 70129\tabularnewline
\hline 
Feb. 2023 & Ge window replaced by sapphire after factor 3 decrease in transmissivity
discovered & 77871\tabularnewline
\hline 
09/11/23 & M. Pedrini\textquoteright s tile AG6831 replaces tiles 404, 405, and
406 & 78687\tabularnewline
\hline 
05/03/24 & Equipped with a 4095 nm long wave pass filter & 79993\tabularnewline
\hline 
23/07/25 & Valley tile AH5251 replaces M. Pedrini's tile AG6831 & 87125\tabularnewline
\hline 
\end{tabular}
\end{table}

\begin{table}
\caption{\label{tab:TIR history}TIR history.}

\centering{}%
\begin{tabular}{|c|>{\centering}p{11cm}|c|}
\hline 
Date & Event & 1st discharge\tabularnewline
\hline 
\hline 
06/06/19 & 1st discharge with TIR & 63060\tabularnewline
\hline 
13/08/19 & Port protection tiles installed in TCV & 63528\tabularnewline
\hline 
12/07/22 & Two long 45º tiles, AF7102 and AF7103, installed & 74982\tabularnewline
\hline 
05/03/24 & Equipped with a 4095 nm long wave pass filter & 79993\tabularnewline
\hline 
20/03/25 & Upper port support installed, for RE studies & 85460\tabularnewline
\hline 
29/04/25 & Lower port camera support rotated by 28º for ELM studies & 85885\tabularnewline
\hline 
23/07/25 & Rooftop tiles AH6513 \& AH6514 (for ELM studies) replace tiles AF7102
\& AF7103 & 87125\tabularnewline
\hline 
\end{tabular}
\end{table}

\subsection{VIR: the vertical infrared system}

VIR is the IR system most used in TCV, as it images the floor of TCV,
where the outer strike point of TCV\textquoteright s single null plasmas
is usually localized. It also views the tilted tiles (Fig. \ref{fig:HIR =000026 VIR view}).
Originally, the VIR system had a Thermosensorik camera (for specification
see Sec. 2.2.1.1 of \cite{Maurizio 2020 PhD thesis}), which was replaced
for an IRCAM Equus2 81k M in March 2017. VIR uses a relay optics with
seven lenses (six of silicon, one of germanium) and a mirror (Sec.
4.3.1 of \cite{Colandrea 2024 PhD thesis}). Table \ref{tab:VIR history}
describes a summary of VIR's history.

\subsection{TIR: the tangential infrared system}

TIR is the newest thermography system in TCV (1st discharge: 63060,
in 2019). Its main view maps the lower tilted tiles and port protection
tiles (Fig. \ref{fig: TIR view}), which can be used to measure the
outer strike point of alternative divertor configurations such as
the super-X, X-point target, and negative triangularity plasmas. In
December 2024, a new upper port view was installed for the TIR, mainly
to measure synchrotron radiation and heat deposition due to REs. The
camera can be rotated into 5 different positions in this view. The
TIR history is summarized in Table \ref{tab:TIR history}. As of discharge
85885, the camera view has been rotated by 28º, to align its vertical
direction with the tilted tiles, allowing smaller frame size and increased
acquisition frequency.

\section{\label{sec:Calibration-to-temperature}Calibration to temperature}

The model used in TCV to relate the IR camera digitized counts $N_{ij}$
at a given pixel $ij$ ($i\in[1,256]$ and $j\in[1,320]$) to the
tile temperature $T_{ij}$ is
\begin{equation}
N_{ij}(T_{ij},\Delta t_{int})=\left[a_{ij}\mathcal{I}(T_{ij})+b_{ij}\right]\Delta t_{int}+c_{ij}\label{eq:Nij}
\end{equation}
where $\mathcal{I}(T)\equiv R_{BB}(T)/(2c)$ (with $c$ being the
light speed and $R_{BB}$ the total photon radiance, given by Eq.
(\ref{eq: R_BB(T)})), $a_{ij}\;[\text{counts sr m}^{3}/\text{ms}]$
is the effective transmittance of the signal, $b_{ij}\;[\text{counts/ms}]$
is a term accounting for dark currents and infrared emission from
objects other than the tiles (e.g. lenses, windows), and $c_{ij}\;[\text{counts}]$
corresponds to the base level of the pixels. More details on the different
effects that compose $a_{ij}$ can be found in Sec. 4.2 of \cite{Colandrea 2024 PhD thesis}.

To obtain the parameters $a_{ij}$, $b_{ij}$, $c_{ij}$, a two-step
procedure is employed.

\subsection{\label{subsec: Calib step 1}Step 1 --- Heated tile}

\begin{figure}
\begin{centering}
\includegraphics[scale=0.55]{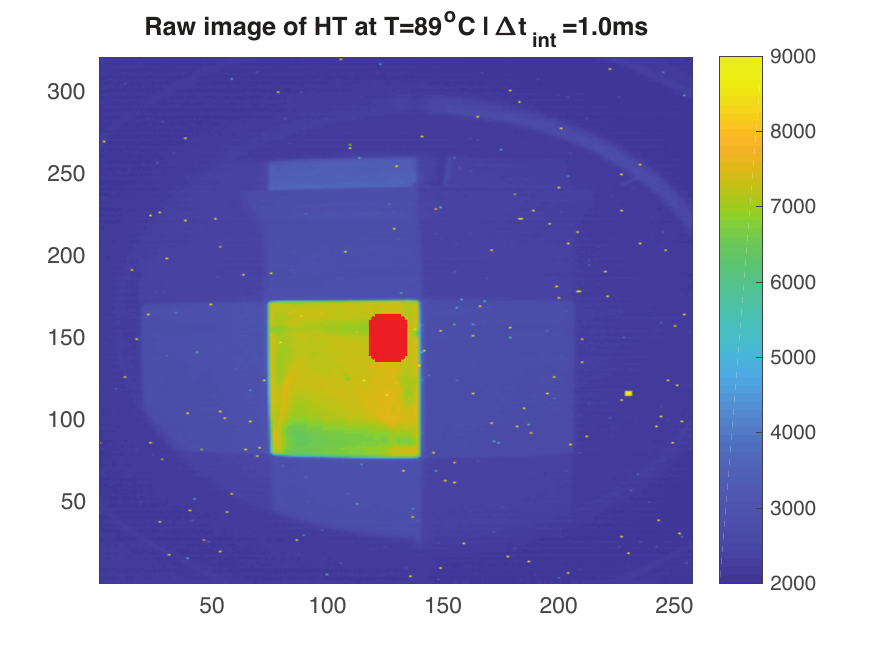}
\par\end{centering}
\caption{\label{fig:Heated-tile}HIR raw data view during the temperature calibration
with TC at $T=89\;\text{ºC}$. Heated tile (HT) in the center, with
the pixels $ij$ used for the calibration in red.}
\end{figure}

\begin{figure}
\subfloat[]{\begin{centering}
\includegraphics[scale=0.55]{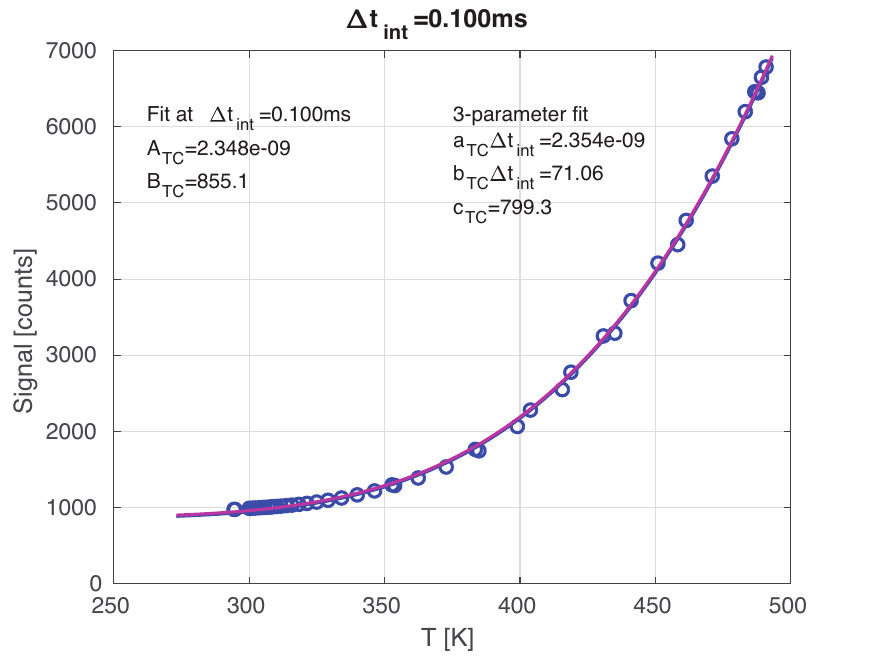}
\par\end{centering}

}
\begin{centering}
\subfloat[]{\begin{centering}
\includegraphics[scale=0.55]{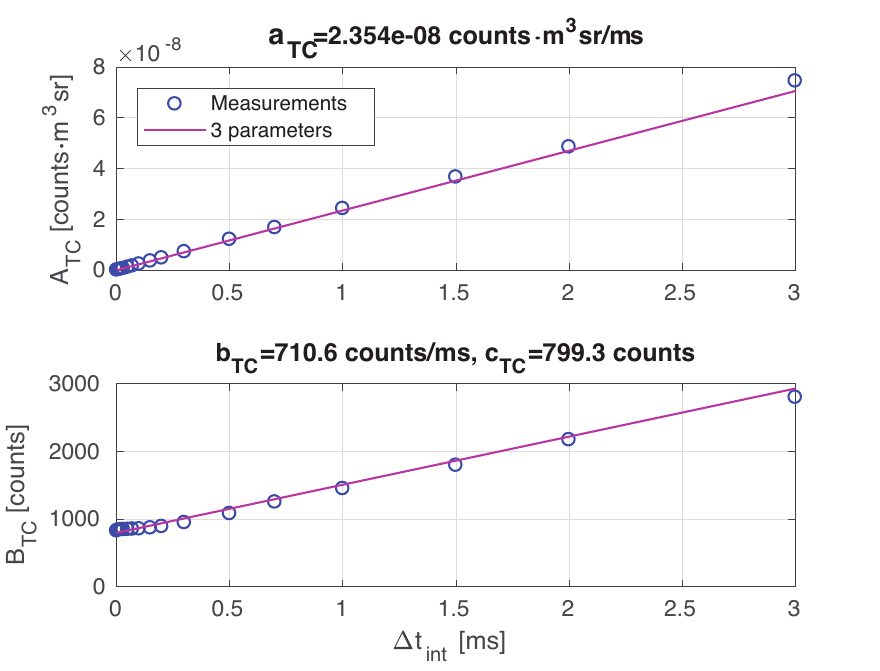}
\par\end{centering}
}
\par\end{centering}
\caption{\label{fig:calib N-vs-T}(a) Raw HIR data vs TC temperature for $\Delta t_{int}=0.1\;\text{ms}$.
(b) Fits of $A_{TC}$ and $B_{TC}$ vs $\Delta t_{int}$. Fitting
Eq. (\ref{eq: Ntc vs Atc Btc}) in Fig. \ref{fig:calib N-vs-T}(a)
gives $A_{TC}$ and $B_{TC}$ at $\Delta t_{int}=0.1\;\text{ms}$.
Doing the same for different integration times yields the data of
Fig. \ref{fig:calib N-vs-T}(b), which can then be fitted by Eq. (\ref{eq: Atc Btc})
to obtain $a_{TC}$, $b_{TC}$, and $c_{TC}$ (known as 3 parameter
fit).}
\end{figure}

A heated tile embedded with a thermocouple (TC) is used to calibrate
the camera signal to the tile temperature (Fig. \ref{fig:Heated-tile}),
\begin{equation}
N_{TC}(T_{TC},\Delta t_{int})=A_{TC}(\Delta t_{int})\mathcal{I}(T_{TC})+B_{TC}(\Delta t_{int})\label{eq: Ntc vs Atc Btc}
\end{equation}
where $N_{TC}$ is the average of the camera signal $N_{ij}$ over
the closest pixels to the TC $(N_{TC}\equiv\left\langle N_{ij}\right\rangle _{TC})$
and
\begin{align}
A_{TC}(\Delta t_{int}) & \equiv a_{TC}\Delta t_{int}\label{eq: Atc Btc}\\
B_{TC}(\Delta t_{int}) & \equiv b_{TC}\Delta t_{int}+c_{TC}\nonumber 
\end{align}
with $a_{TC}\equiv\left\langle a_{ij}\right\rangle _{TC},$ $b_{TC}\equiv\left\langle b_{ij}\right\rangle _{TC},$
and $c_{TC}\equiv\left\langle c_{ij}\right\rangle _{TC}.$ Graphs
of $N_{TC}(T_{TC},\Delta t_{int})$ vs $T_{TC}$ are obtained for
several integration times (Fig. \ref{fig:calib N-vs-T}). Fitting
them yields $A_{TC}(\Delta t_{int})$ and $B_{TC}(\Delta t_{int})$,
which are then used to evaluate $a_{TC}$, $b_{TC}$, and $c_{TC}$
from fits of Eq. (\ref{eq: Atc Btc}).

\subsection{\label{subsec: Calib step 2}Step 2 --- Non-uniformity correction}

Step 1 obtains the calibration parameters at the thermocouple position.
To evaluate $a_{ij},$ $b_{ij},$ and $c_{ij}$ for all other pixels,
a non-uniformity correction is needed. For this, IR images of a blackbody
(e.g. a piece of paper in front of the camera) at uniform temperature
$T_{0}$ are taken at different integration times. The signal $N_{ij}(T_{0},\Delta t_{int})$
at each pixel can then be related to

\begin{equation}
N_{ij}(T_{0},\Delta t_{int})=a'_{ij}(T_{0})\Delta t_{int}+c_{ij}\label{eq:Nij NUC}
\end{equation}
where $a'_{ij}(T_{0})\equiv a_{ij}\mathcal{I}(T_{0})+b_{ij}$ (from
Eq. (\ref{eq:Nij})). Eq. (\ref{eq:Nij NUC}) is then fitted at different
integration times, yielding $a'_{ij}$ and $c_{ij}$ for all pixels
$i\in[1,256]$ and $j\in[1,320]$. The gain and offset indices for
each pixel are then evaluated,
\begin{align*}
g_{ij} & =\frac{a'_{ij}}{a'_{TC}}\\
o_{ij} & =\frac{c_{ij}}{c_{TC}}
\end{align*}
where $a'_{TC}$ is the averaged $a'_{ij}$ over the pixels around
the thermocouple used in step 1. This results in the coefficients
\begin{align*}
a_{ij} & =g_{ij}a_{TC}\\
b_{ij} & =g_{ij}b_{TC}\\
c_{ij} & =o_{ij}c_{TC}
\end{align*}
which are then used in Eq. (\ref{eq:Nij}) for each IR camera during
TCV shots to translate the raw signal to surface temperature. The
matrices of $a_{ij}$, $b_{ij}$, and $c_{ij}$ are shown in Fig.
.

\begin{figure}
\begin{centering}
\includegraphics[scale=0.55]{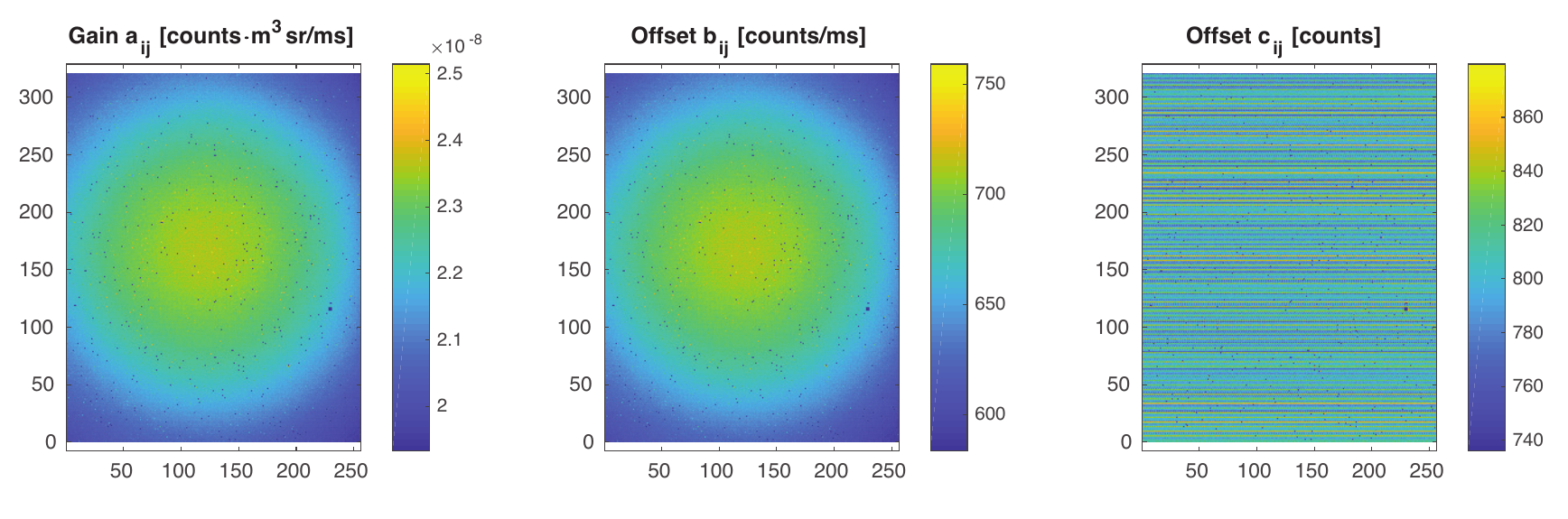}
\par\end{centering}
\caption{\label{fig: aij, bij, cij} $a_{ij}$, $b_{ij}$, and $c_{ij}$ matrices
for the HIR calibration shown in Fig. \ref{fig:calib N-vs-T}.}
\end{figure}

\section{\label{sec:Heat-flux-estimation}Heat flux estimation}

\subsection{\label{subsec: THEODOR}From temperature to heat flux with THEODOR}

The heat flux impinging perpendicularly to the tile is evaluated through
Fourier's law,
\begin{equation}
\vec{q}_{\perp}=-k\nabla T\label{eq: Fourier's law}
\end{equation}
where $k(T)$ is the tile conductivity and $T(x,y,t)$ is the tile
temperature along a line of interest, with $x$ being the coordinate
along the heat flux profile and $y$ the coordinate that goes from
the surface to inside the tile. To obtain the temperature distribution
in the tile, the heat diffusion equation is used,

\begin{equation}
\rho c_{p}\frac{\partial T}{\partial t}=\nabla\cdot(k\nabla T)\label{eq: heat diff}
\end{equation}
where $\rho=cte$ is the tile mass density and $c_{p}(T)$ is the
tile specific heat. $\rho$, $c_{p}$, and $k$ are known. As it will
be shown in Sec. \ref{subsec: NPL}, the $c_{p}$, and $k$ depend
on the tile temperature, which complicates the solution of Eq. (\ref{eq: heat diff}).
However, introducing the heat flux potential 
\begin{equation}
U(T)\equiv\int_{0}^{T}k(T)dT,\label{eq: U def}
\end{equation}
Eq. (\ref{eq: heat diff}) can be rewritten as (see Appendix B of
\cite{Sieglin 2014 PhD thesis})
\begin{equation}
\frac{\partial U}{\partial t}=D\nabla^{2}U\label{eq: dU/dt}
\end{equation}
where the tile heat diffusivity is defined as
\begin{equation}
D(T)\equiv\frac{k(T)}{\rho c_{p}(T)}\label{eq: D=00003Dk/pc}
\end{equation}
Eq. (\ref{eq: dU/dt}) is solved by the code THEODOR (\textbf{Th}ermal
\textbf{E}nergy \textbf{O}nto \textbf{D}ivert\textbf{or}) \cite{Herrmann 2001 THEODOR}
to obtain $U$ as a function of time. 

The following parametrization is used for the conductivity, 

\begin{equation}
k(T)=a_{k}+b_{k}\left(1+\frac{T}{T_{0,k}}\right)^{-2}\label{eq: cond}
\end{equation}
where $a_{k}$, $b_{k}$ and $T_{0,k}$ are parameters to be determined
with the available data. In TCV, $a_{k}=0$ was obtained (Sec. \ref{subsec: NPL}).
The same functional dependence is used for the tile diffusivity,
\begin{equation}
D(T)=a_{D}+b_{D}\left(1+\frac{T}{T_{0,D}}\right)^{-2}\label{eq: diff}
\end{equation}
Integrating Eq. (\ref{eq: U def}) using $k(T)$ from Eq. (\ref{eq: cond})
yields
\[
U(T)=a_{k}T+b_{k}\frac{T}{1+T/T_{0,k}}
\]
Inverting this equation results in the tile temperature distribution,
which is then used to obtain the heat flux perpendicular to the tile,
$\vec{q}_{\perp}=-k\nabla T$ (Eq. (\ref{eq: Fourier's law})).

\subsection{Boundary conditions and surface layer heat transmission factor}

For IR in TCV, a python version of the THEODOR code is used, which
solves the differential equation (\ref{eq: dU/dt}) implicitly. Three
boundary conditions are used for this THEODOR version. Two of them
are
\[
\begin{cases}
\left.\frac{\partial T}{\partial x}\right|_{\text{side}}=0\\
q_{\perp}(x,d_{tile},t)=0
\end{cases}
\]
The first equation means that THEODOR presumes a vanishing temperature
derivative at the edges of the heat flux profile. The second equation
assumes that the heat flux at the bottom of the tile is zero.

Tile erosion and particle deposition can form surface layers (SLs)
that cause an overestimation of the actual tile temperature \cite{Herrmann 2001 THEODOR,Andrew 2005 alpha,Devaux 2011 alpha}.
In TCV, the main SL components are believed to be made of carbon (wall
material), deuterium (injected gas), and boron oxides (due to boronizations).
To account for this SL and avoid false negative heat fluxes, a surface
layer heat transmission factor is introduced,

\[
\alpha_{top}=\frac{k_{layer}}{d_{layer}}
\]
where $k_{layer}$ and $d_{layer}$ are respectively the conductivity
and thickness of the layer. Differently from previous IDL versions,
the Python THEODOR does not use use the boundary condition $T(x,0,t)=T_{layer}(x,t)-\frac{q_{\perp}(x,0,t-\Delta t)}{\alpha_{top}}$
(as it was the case in Eq. (4.11) of \cite{Colandrea 2024 PhD thesis}).
Instead, an actual layer is modeled on top of the tile using $\alpha_{top}$
as input. The SL thickness is assumed to be of $d_{layer}=10\;\mu\text{m}$,
although in reality this value could change depending on the tile
exposition to the plasma. The SL conductivity is then obtained with
$k_{layer}=\alpha_{top}d_{layer}$. The SL mass density times heat
capacity, $\rho_{layer}c_{p,layer}$, is supposed to be 100 times
smaller than the top of the tile just below the SL, which is then
used to estimate the SL diffusivity, $D_{layer}=k_{layer}/(\rho_{layer}c_{p,layer})$.

Values for $\alpha_{top}$ in TCV range from $10\;\text{kW/m}^{2}/\text{K}$
in tile regions with very thick surface layers to $160\;\text{kW}/\text{m}^{2}/\text{K}$
in clean tiles. $\alpha_{top}$ is estimated as the maximum value
which avoids negative heat fluxes.

\subsection{Power and parallel heat flux}

The heat flux perpendicular to the tile $q_{\perp}$ is obtained with
THEODOR. It can be related to the plasma heat flux parallel to the
magnetic field through the equation
\[
q_{\perp}=q_{B,\parallel}\sin\alpha+q_{B,\perp}\cos\alpha+q_{bg}
\]
where $\alpha$ is the grazing angle between the plasma and the tile
(not to be confused with $\alpha_{top}$, the surface layer heat transmission
factor), $q_{B,||}$ is the plasma heat flux parallel to the magnetic
field, $q_{B,\perp}$ is the plasma heat flux perpendicular to the
magnetic field and $q_{bg}$ is the background heat flux due to radiation.
Assuming that for diverted plasmas $q_{B,||}\sin\alpha\gg q_{B,\perp}\cos\alpha$
near the strike points, the heat flux parallel to $B$ can be estimated
as
\begin{equation}
q_{\perp}=q_{\parallel}\sin\alpha+q_{bg}\implies q_{||}=\left(q_{\perp}-q_{bg}\right)/\sin\alpha\label{eq: qpar def}
\end{equation}
where for simplicity $q_{||}\equiv q_{B,||}$.

Some tiles in TCV are not toroidally symmetric (e.g. all inner wall
tiles seen by HIR, the VIR valley tile, and the TIR tiles). To evaluate
the perpendicular heat flux on a toroidally symmetric tile, a correction
in the grazing angle is made, yielding 
\begin{equation}
q_{\perp,\text{sym}}=q_{||}\sin\alpha_{\text{sym}}+q_{bg}\label{eq: qperp sym def}
\end{equation}
If the tile is toroidally symmetric (i.e. if $\alpha_{\text{sym}}=\alpha$,
as for VIR's flat tiles), then $q_{\perp,\text{sym}}=q_{\perp}$.

The power deposited on TCV's walls is obtained by integrating $q_{\perp,\text{sym}}$, 

\begin{equation}
P=\int_{s_{min}}^{s_{max}}\int_{0}^{2\pi}q_{\perp,\text{sym}}Rd\varphi ds=2\pi\int_{s_{min}}^{s_{max}}q_{\perp,\text{sym}}Rds\label{eq: power def}
\end{equation}
where $s=z$ for the inner wall, $s=R$ for the floor, and $s=\sqrt{R^{2}+z^{2}}$
for the tilted tiles.

It is also possible to project the parallel heat flux upstream, by
accounting for the total flux expansion from the upstream to the target,
\[
q_{||,u}=f_{x,tot}q_{||}\approx\frac{B_{u}}{B_{t}}q_{||}
\]
where $B_{u}$ is the magnetic field upstream (on the outer midplane)
and $B_{t}$ is the magnetic field at the target. It is common to
show the $q_{||,u}$ profile as a function of the upstream coordinate
$R-R_{sep}$ where $R$ here is the major radius position upstream
and $R_{sep}$ is the major radius of the separatrix. 

\section{\label{sec:Windows}Windows}

TCV's IR Equus cameras request that the windows which separate the
tokamak to the camera have a high transmissivity in the range $3.7\;\mu\text{m}$
to $4.8\;\mu\text{m}$ (Table \ref{tab:Gen-Info}), which can be obtained
for example with zinc selenide (ZnSl), germanium (Ge) or sapphire
($\text{Al}_{2}\text{O}_{3}$). Currently all TCV IR windows are made
of sapphire with an anti-reflection coating. Previously used germanium
windows cracked (HIR lower port, Table \ref{tab:HIR history}) or
had their transmission to decay by a factor 3 with time (VIR, Table
\ref{tab:VIR history}). Zinc selenide are more fragile and are therefore
avoided.

\section{\label{sec:Tiles}Tiles}

The tiles, together with the infrared cameras, compose the core of
the IR system. Heat fluxes are inferred from the tile surface temperatures
(Sec. \ref{subsec: THEODOR}). TCV's tiles are made of polycrystalline
graphite, which has an emissivity in the range of 0.7--0.9 (and therefore
a reflectivity in the range 0.1-0.3).

\subsection{\label{subsec: NPL}Tile mass density, specific heat, diffusivity
and conductivity}

\begin{figure}
\begin{centering}
\subfloat[]{\begin{centering}
\includegraphics[scale=0.5]{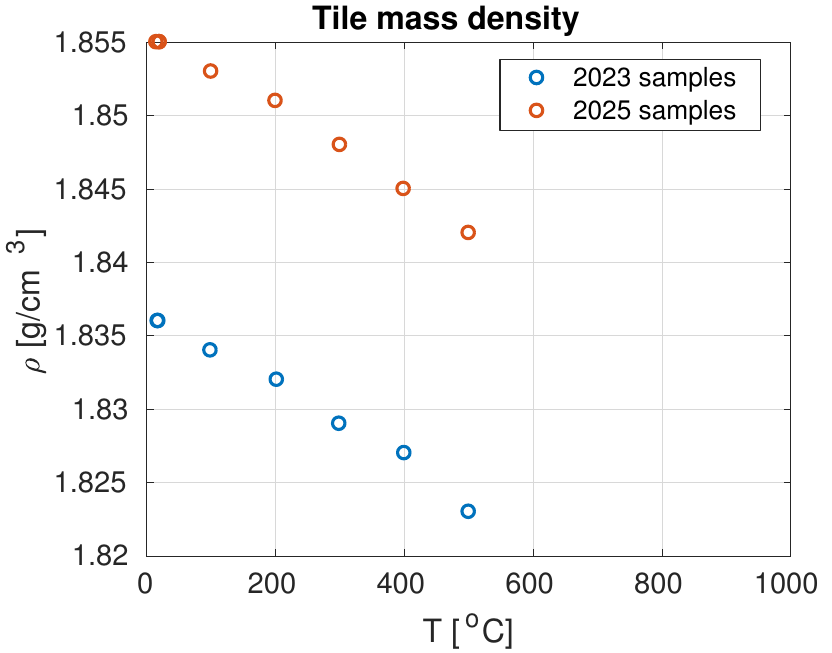}
\par\end{centering}
}\subfloat[]{\begin{centering}
\includegraphics[scale=0.5]{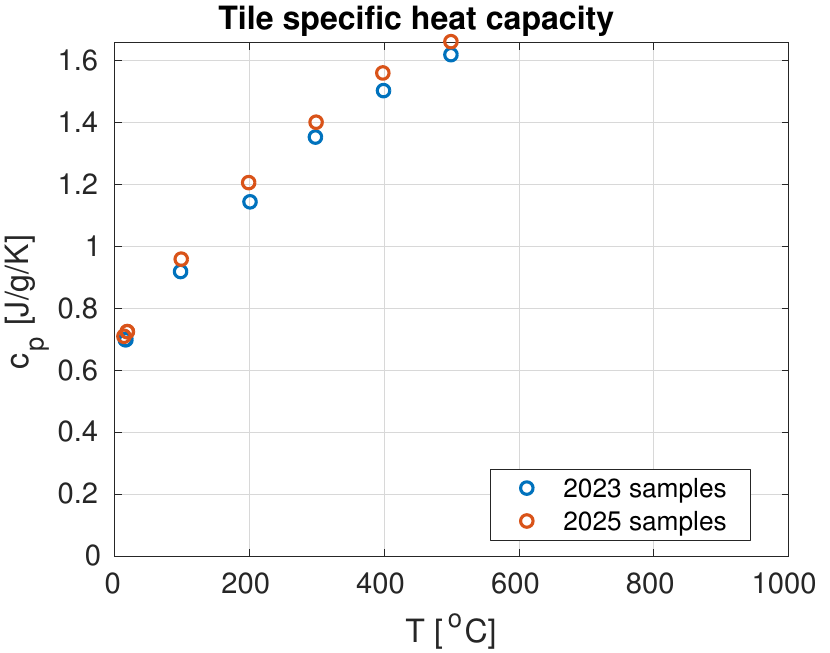}
\par\end{centering}
}
\par\end{centering}
\begin{centering}
\subfloat[]{\begin{centering}
\includegraphics[scale=0.5]{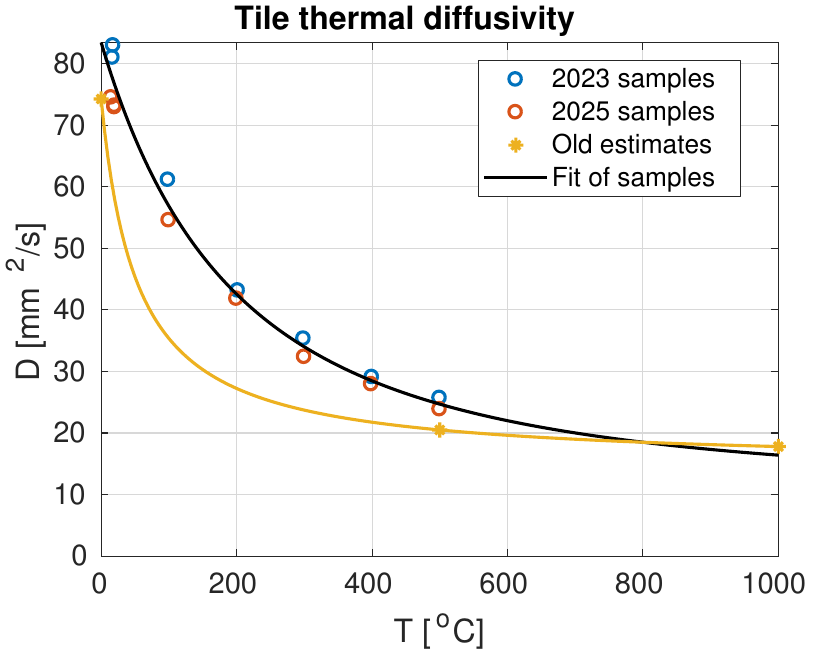}
\par\end{centering}
}\subfloat[]{\begin{centering}
\includegraphics[scale=0.5]{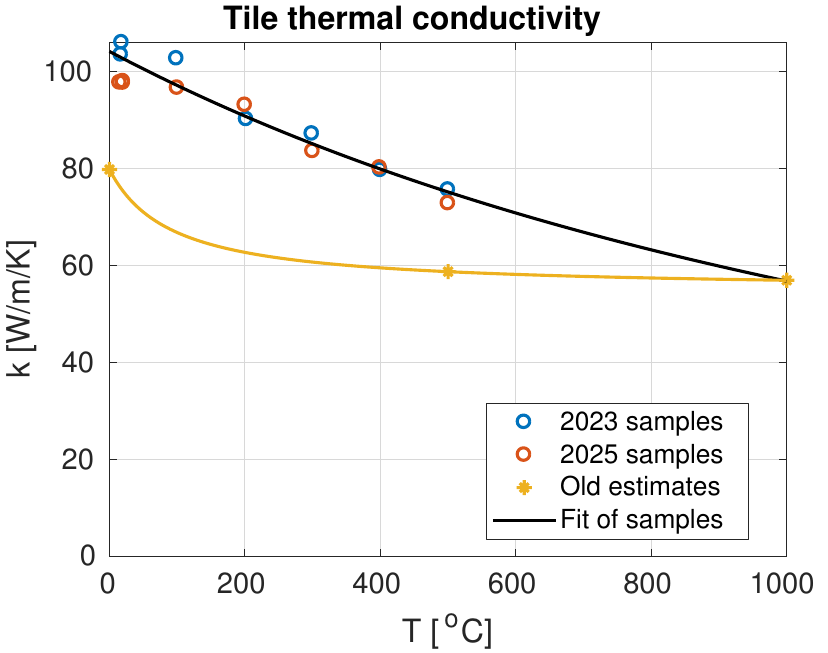}
\par\end{centering}
}
\par\end{centering}
\caption{\label{fig:NPL}TCV's graphite tile properties measured by NPL. (a)
Mass density. (b) Specific heat capacity. (c) Diffusivity. (d) Conductivity.
Blue and circles: data from 2023 samples; orange circles: 2025 samples.
Black curves: fit using the functional dependency expected by THEODOR
($f(T)=a+b(1+T/T_{0})^{-2}),$ yielding Eq. (\ref{eq: D fit}) and
(\ref{eq: k fit}). Yellow stars and curves: old estimates used in
TCV for the tile diffusivity and conductivity (Eq. (\ref{eq: old diff})
and (\ref{eq: old cond})).}
\end{figure}

Using IR to infer plasma heat fluxes impinging on the tiles requires
knowledge about the tile conductivity and diffusivity (Eq. (\ref{eq: Fourier's law})
and (\ref{eq: dU/dt})). For this reason, thermal properties of TCV
tile samples were measured in 2023 and 2025 by S. Smith, P. Mildeova,
Y. Mouasher, A. Awad, and L. Orkney from the National Physics Laboratory
(NPL) of the United Kingdom. The results are shown in Fig. \ref{fig:NPL}
and in the Tables \ref{tab: NPL 2023}, and \ref{tab: NPL 2025} from
the Appendix \ref{sec:Appendix:NPL tables}.

The samples' mass densities were measured at room temperature with
a calibrated Mettler AE240 five-figure electronic balance, obtaining
$\rho=1.836\;\text{g}/\text{cm}^{3}$ for the 2023 sample and $\rho=1.855\;\text{g}/\text{cm}^{3}$
for the 2025 one. Thermal expansion measurements (not shown) were
then performed with a Linseis twin push-rod alumina dilatometer up
to a maximum temperature of 1020 ºC. The thermal expansion measurements
were used to evaluate the mass density at different temperatures.
The mass density variation with temperature was negligible: it decreased
by less than 1\% in the range of 20 ºC to 500 ºC (Fig \ref{fig:NPL}(a)).

The specific heat capacity was obtained with a TA Instruments twin-pan
differential scanning calorimeter (DSC). The test-piece mass was determined
with a calibrated Sartorius Secura 225D-1S electronic balance to a
resolution of $0.00001\;\text{g}$. Three heating runs were performed
on each sample, from 0 ºC to 430 ºC. It is seen in Fig. \ref{fig:NPL}(b)
that the specific heat increases with temperature, going from $0.7\;\text{J/g/K}$
at 20 ºC to $1.6\;\text{J/g/K}$ at $500\;\text{ºC}$ (the results
above $430\;\text{ºC}$ was \emph{extrapolated}).

The thermal diffusivity of the samples was determined via a laser
flash analysis (LFA) using the instrument Netzsch LFA 427. The tile
diffusivity decreases with the functional dependency expected by THEODOR
(Eq. (\ref{eq: diff})), ranging from around $80\;\text{mm}^{2}/\text{s}$
at $T=0\;\text{º}\text{C}$ to $25\;\text{mm}^{2}/\text{s}$ at $500\;\text{º}\text{C}$.
Fitting the data from the three samples yielded (black curve in Fig.
\ref{fig:NPL}(c))
\begin{equation}
\text{\fbox{\ensuremath{D_{fit}(T)=\left[10.6(2.7)\;+72.9(2.9)\left(1+\frac{T}{392(56)\;\text{ºC}}\right)^{-2}\right]\text{mm}^{2}/\text{s}}}}\label{eq: D fit}
\end{equation}
where the values in parentheses ((2.7), (2.9), (56)) are the uncertainties
in the parameters. Using the standard deviation between the 2023 and
2025 samples as uncertainty yielded a chi-squared of $\chi^{2}=7$,
in good agreement with the number of degrees of freedom, $\text{DF}=12$.
Before the NPL measurements, the old estimate used in TCV for the
tile diffusivity was (yellow curve in Fig. \ref{fig:NPL}(c))
\begin{equation}
D_{old}(T)=\left[15\;+60\left(1+\frac{T}{53\;\text{ºC}}\right)^{-2}\right]\text{mm}^{2}/\text{s}\label{eq: old diff}
\end{equation}

Finally, the sample conductivity data was obtained by multiplying
the mass density, specific heat, and diffusivity $k=\rho c_{p}D$
(as in Eq. (\ref{eq: D=00003Dk/pc})). A least-squares fit yielded
(black curve in Fig. \ref{fig:NPL}(d))

\begin{equation}
\text{\fbox{\ensuremath{k_{fit}(T)=104.2(1.7)\left(1+\frac{T}{2.82(0.19)\times10^{3}\;\text{ºC}}\right)^{-2}\frac{\text{W}}{\text{m}\;\text{K}}}}}\label{eq: k fit}
\end{equation}
$a_{D}$ in Eq. (\ref{eq: D=00003Dk/pc}) was set to zero since keeping
it still yielded a value compatible to 0 but with much higher uncertainties
in the parameters, which is a sign of over-parametrization. The chi-squared
was $\chi^{2}=9$, again in good agreement with the degrees of freedom,
$\text{DF}=13$. The tile conductivity decreases from 100 W/m/K at
0ºC to 75 W/m/K at 500 ºC. The old values of thermal conductivity
(yellow curve in Fig. \ref{fig:NPL}(d)) used for TCV's IR were smaller
than the NPL results by 25\% to 40\% for temperatures up to 500 ºC,

\begin{equation}
k_{old}(T)=\left[55+25\left(1+\frac{T}{92\;\text{ºC}}\right)^{-2}\right]\frac{\text{W}}{\text{m}\;\text{K}}\label{eq: old cond}
\end{equation}

As of January 2026, the tile diffusivity and the conductivity given
by Eq. (\ref{eq: D fit}) and (\ref{eq: k fit}) are used for IR in
TCV to evaluate heat fluxes with THEODOR. The effect on the estimated
heat fluxes was small --- less then 15\% increase in the peak heat
flux when compared to the old estimated thermal properties, with the
difference going to 0 far from the peak.

\subsection{\label{subsec:VIR-valley-tile}VIR valley tile}

\begin{figure}
\begin{centering}
\subfloat[]{\begin{centering}
\includegraphics[scale=0.5]{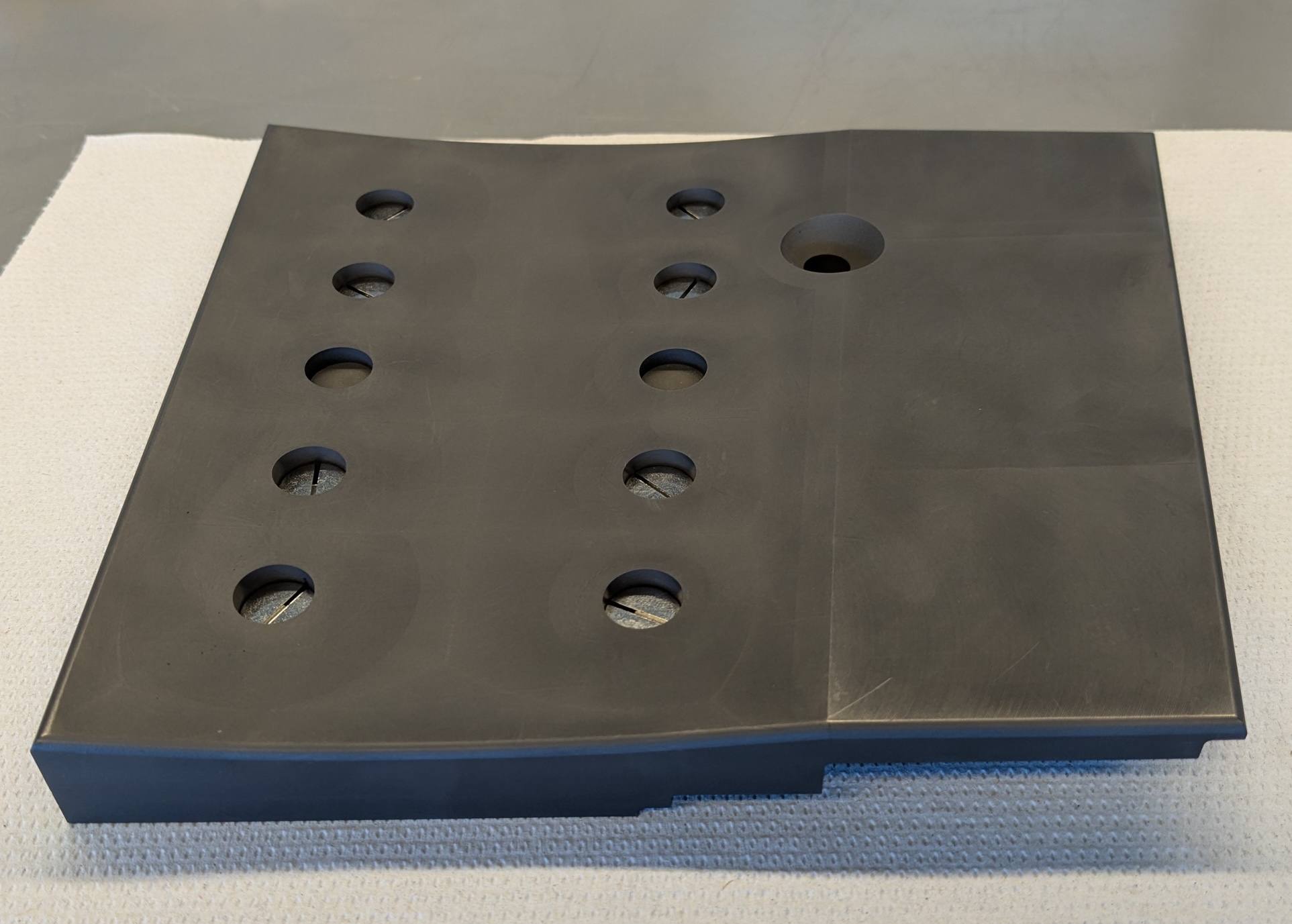}
\par\end{centering}
}\subfloat[]{\begin{centering}
\includegraphics[scale=0.16]{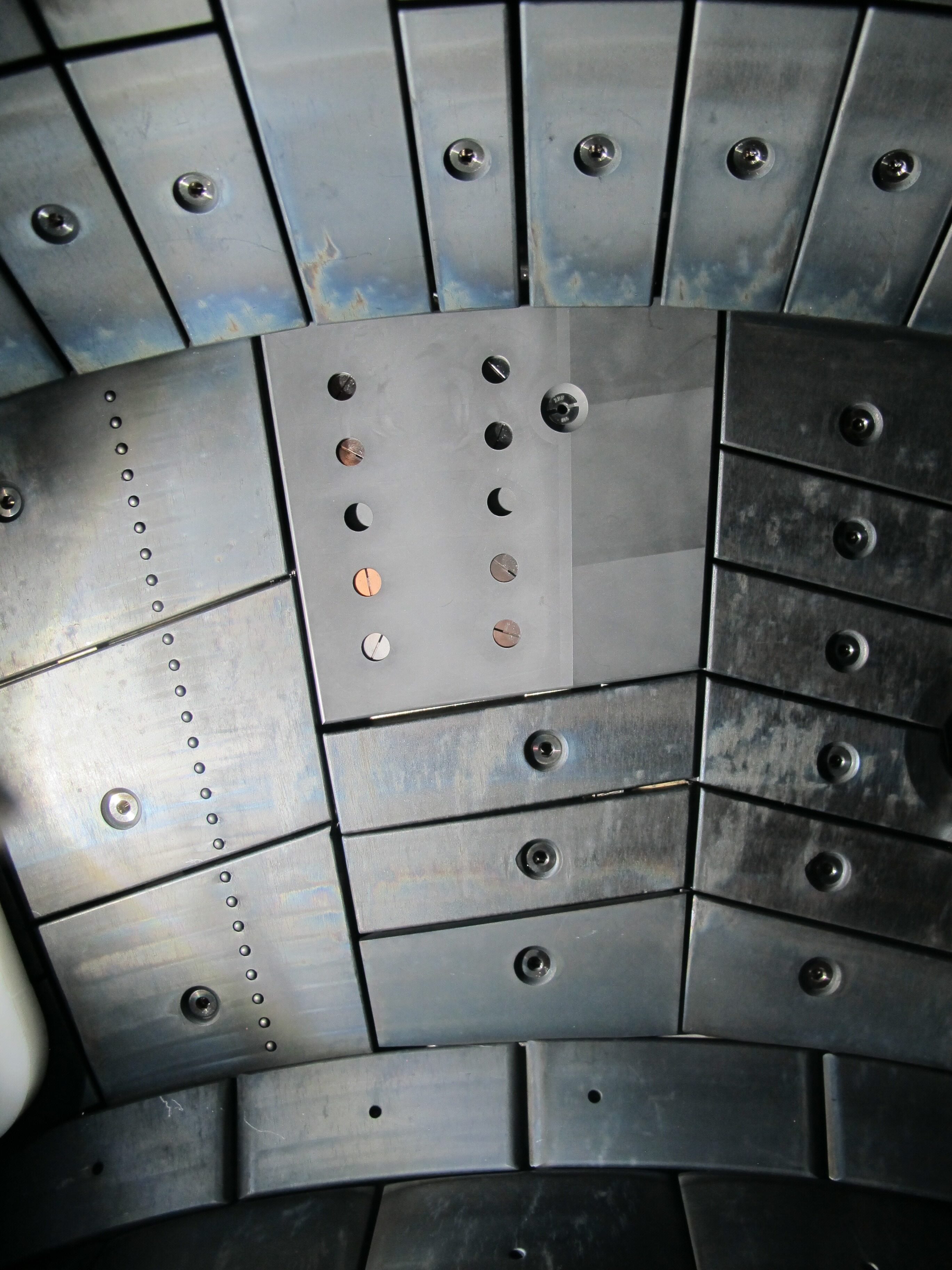}
\par\end{centering}
}
\par\end{centering}
\caption{\label{fig:VIR-valley-tile}(a) VIR valley tile outside TCV. (b) TCV
floor with the VIR valley tile, also showing tiles with Langmuir probes
on the left. Photo (b) by Frédéric Dolizy and Luc Morier-Genoud.}
\end{figure}

A new valley tile has been commissioned for the VIR system in 2025
(Fig. \ref{fig:VIR-valley-tile}). The main goal of the tile is to
allow the camera operate at high acquisition frequencies. This is
achieved by two means, both which allow decreasing the camera integration
time and therefore increasing the acquisition frequency ($f_{acq}=1/\Delta t_{frame}$,
with $\Delta t_{frame}\approx\Delta t_{int}+\Delta t_{read}$ given
by Eq. (\ref{eq: dt frame final})). Firstly, the tile has been embedded
with a heating element, which can be used to heat up the tile just
before the shot. Due to the highly non-linear dependence of the infrared
radiation with the tile temperature, heating it allows significant
increase in the camera signal. 

Secondly, the tile has parts with an inclination of 6º to increase
the grazing angle, which also leads to a higher tile temperature,
again allowing lower integration times. The reasoning is summarized
in the following equation:

\[
\Delta T\propto q_{\perp}\sqrt{\Delta t}\approx\sin(\alpha)q_{\parallel}\sqrt{\Delta t}\propto\sin\alpha\approx\alpha
\]
The variation in the tile surface temperature $\Delta T$ due to a
heat flux $q_{\perp}$ perpendicular to the tile applied during an
interval $\Delta t$ is proportional $q_{\perp}\sqrt{\Delta t}$ \cite{Carlslaw conduction of heat}.
The heat flux perpendicular to the tile $q_{\perp}$ is then approximately
connected to the heat flux parallel to the magnetic fields through
$q_{\perp}\approx q_{\parallel}\sin\alpha$. Therefore, increasing
the grazing angle $\alpha$ increases $\Delta T$ almost linearly
(since $\alpha\ll1\implies\sin\alpha\approx\alpha$).

Tilting the tile in the toroidal direction by an angle $\alpha_{ti}$
then increases the temperature difference by
\[
\frac{\Delta T}{\Delta T_{0}}\propto\frac{\alpha_{0}+\alpha_{ti}}{\alpha_{0}}
\]
where $\alpha_{0}$ is the grazing angle between the plasma and a
flat floor tile. It was chosen to incline the tile to $\alpha_{ti}=6\text{º}$,
such than the increase in the temperature difference would be of a
factor 3 for an initial grazing angle of $\alpha_{0}=3\text{º}$,
$(3\text{º}+6\text{º})/3\text{º}=3$.

The VIR tile was designed as a valley instead of a rooftop to avoid
shadowing Langmuir probes present in neighboring tiles (small dots
in Fig. \ref{fig:VIR-valley-tile}(b)). Additional holes with screws
were added to allow the correction of vibration of the IR images even
in reduced frame mode. However, it was found that the tile temperature
can be modified in the vicinity of these screws, hindering the IR
heat flux analysis. Therefore, if in the future a new tile is manufactured,
it is recommended to not add these extra holes.

\subsection{\label{subsec:TIR-rooftop-tiles}TIR rooftop tiles}

\begin{figure}
\begin{centering}
\subfloat[]{\begin{centering}
\includegraphics[viewport=0bp -84.3212bp 281bp 126bp,scale=0.6]{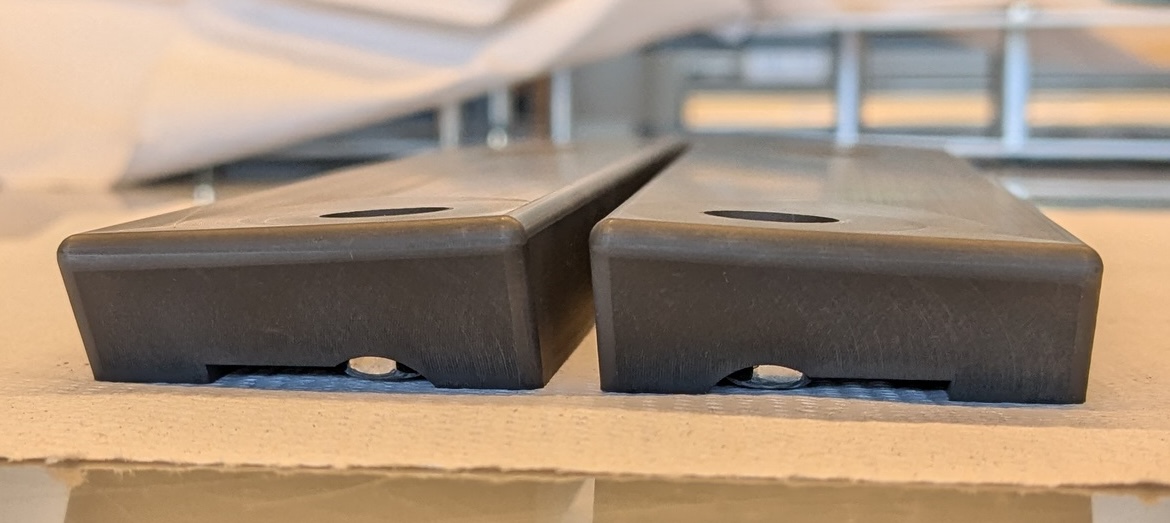}
\par\end{centering}
}\subfloat[]{\begin{centering}
\includegraphics[scale=0.18]{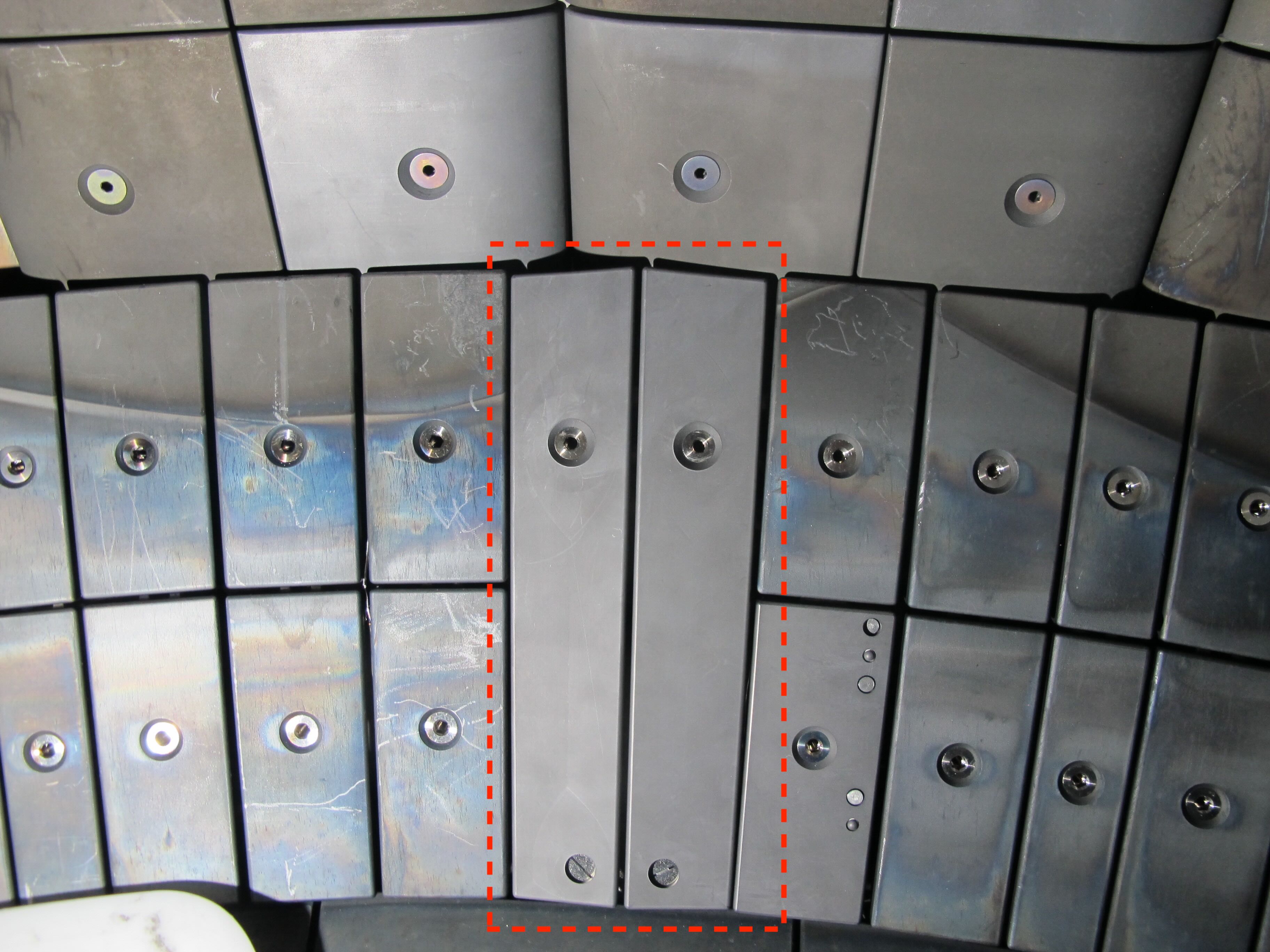}
\par\end{centering}
}
\par\end{centering}
\caption{\label{fig:TIR-rooftop-tile}TIR rooftop tiles outside (a) and inside
(b) TCV. Photo (b) by Frédéric Dolizy and Luc Morier-Genoud.}
\end{figure}

One pair of TIR rooftop tiles has been designed in 2025 (Fig. \ref{fig:TIR-rooftop-tile})
to increase the camera signal by increasing the grazing angle between
the plasma and the tile. The tile was designed with the following
surface equation 

\begin{equation}
z_{t}(y)=\frac{y_{t}^{2}}{2R_{0}}\sin\beta+Ky_{t}\cos\beta\label{eq: zt TIR rooftop}
\end{equation}
where $R_{0}=970\;\text{mm}$ is the major radius at which the tilted
tile starts, $\beta=50\text{º}$ is the angle between the tile and
the floor, $y_{t}$ is the direction of the tile width ($y_{t}\in[-41.9,41.9]\;\text{mm}$
in the lowest part of the tile and $y_{t}\in[-48.8,48.8]\;\text{mm}$
at the highest), $z_{t}$ is the tile height in its referential and
$K=0.0812$ is a constant calculated to transform an initial grazing
angle of $\alpha=4\text{º}$ into $\alpha=7\text{º}$ by the addition
of the rooftop. The derivation of Eq. (\ref{eq: zt TIR rooftop})
is shown in Appendix \ref{sec:Appendix-C TIR rooftop}.

One extra hole with a screw was added to the bottom of each of the
tiles to allow for the correction of vibration of the IR images. However,
it was found that this screw removed significant space for the TIR
line of interest used for the heat flux calculation. Therefore, if
in the future a new TIR rooftop tile is commissioned, it is recommended
to not add this extra screw.

\section{\label{sec:IR-plasma-radiation}Filtering infrared plasma emission}

Infrared plasma radiation (IR-PR) can distort the tile temperature
measurements obtained with infrared thermography, especially at low
plasma temperature and high plasma density. IR-PR can occur in three
forms: bound-bound emission (atoms excited by electrons or charge
exchange, emitting photons known as line emission), free-bound emission
(free electron recombining with an atom and emitting the excess kinetic
energy) or free-free emission (photons created during de-acceleration
of free electrons, known as bremsstrahlung). 

\begin{figure}
\begin{centering}
\includegraphics[scale=0.65]{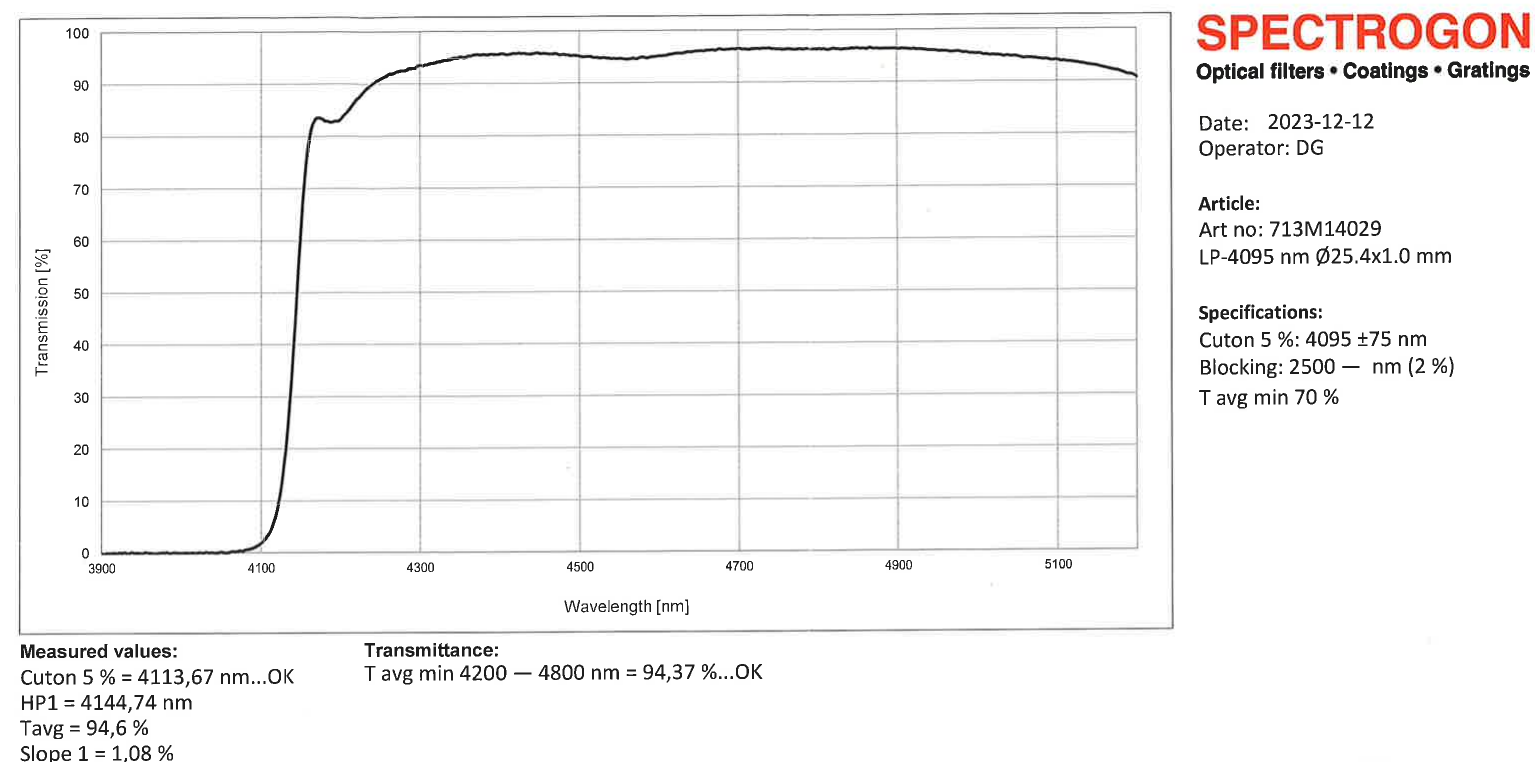}
\par\end{centering}
\caption{\label{fig:IR filter 4095nm}Transmittance of the SPECTROGON LP-4095nm
filter used for the VIR and TIR cameras (and similar one for HIR).}
\end{figure}

According to the NIST LIBS database, the strongest atomic line emission
in the range of TCV's IRCAM Equus cameras ($3.7\;\mu\text{m}<\lambda<4.8\;\mu\text{m}$)
is the deuterium line $D_{5\rightarrow4}$, with a wavelength of $\lambda=4051\;\text{nm}.$
To avoid measuring this line emission, two 4095 nm long wave pass
filters were installed for the VIR and TIR systems (transmissivity
in Fig. \ref{fig:IR filter 4095nm}), whereas HIR already had a 4090
nm filter. Applying this filter also reduces continuous IR-PR, which
is more energetic for lower wavelengths.

\begin{figure}
\begin{centering}
\subfloat[]{\begin{centering}
\includegraphics[scale=0.58]{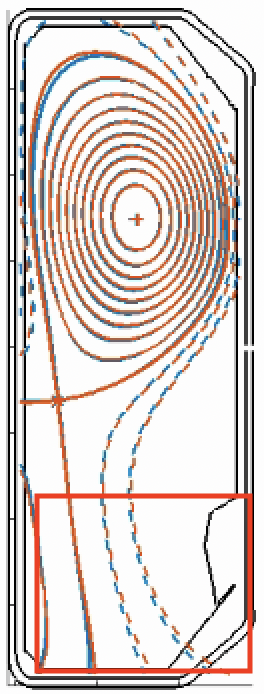}
\par\end{centering}
}\subfloat[]{\begin{centering}
\includegraphics[scale=0.55]{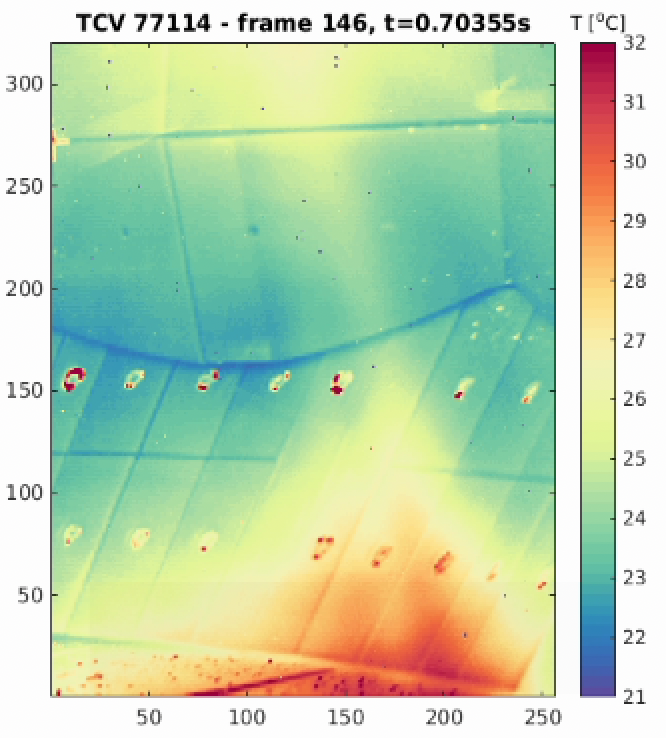}
\par\end{centering}
}\subfloat[]{\begin{centering}
\includegraphics[scale=0.55]{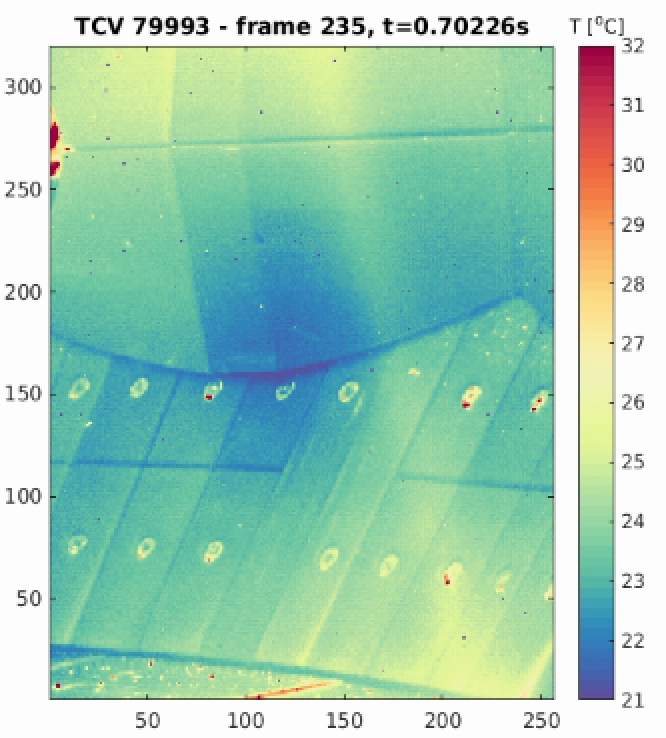}
\par\end{centering}
}
\par\end{centering}
\centering{}\caption{\label{fig:TIR filtered}(a) Magnetic reconstruction of the TCV standard
shots 77114 (blue) and 79993 (orange) at 0.7 s, showing the TIR field
of view (red). (b) Unfiltered TIR view. (c) TIR view after putting
the 4095 nm filter.}
\end{figure}

Fig. \ref{fig:TIR filtered} shows the TIR view before and after mounting
the filter for two identically programmed TCV standard shots, which
are long-legged, ohmic, low temperature discharges (Thomson scattering
pedestal electron temperature and density in the range $T_{e,ped}=10\text{\textendash}20\;\text{eV}$
and $n_{e,ped}=1\text{\textendash}2\times10^{19}\;\text{m}^{3}$ at
$t=0.7\;\text{s}$). A significant reduction in the IR plasma emission
is seen with the filter. Nevertheless, the plasma leg still can be
seen in the filtered image (Fig. \ref{fig:TIR filtered}(b)). After
removing $\lambda<4095\;\text{nm}$, the strongest atomic line emission
is $D_{7\rightarrow5}$, at $\lambda=4654\;\text{nm}$ (carbon line
emissions are negligible in the range $3.7\;\mu\text{m}<\lambda<4.8\;\mu\text{m}$
according to the NIST LIBS database). However, the $D_{7\rightarrow5}$
intensity is 4 times weaker than $D_{5\rightarrow4}$. This, together
with SOLPS-ITER simulations (Sec. 4.5 of \cite{Colandrea 2024 PhD thesis}),
suggest that, after removing $\lambda<4095\;\text{nm}$, free-bound
and free-bound emissions should dominate the measured IR-PR for TCV's
IRCAM Equus cameras. 

\section{\label{sec:VIR-vs-TIR}Comparing VIR and TIR heat fluxes}

\begin{figure}
\begin{centering}
\includegraphics[scale=0.4]{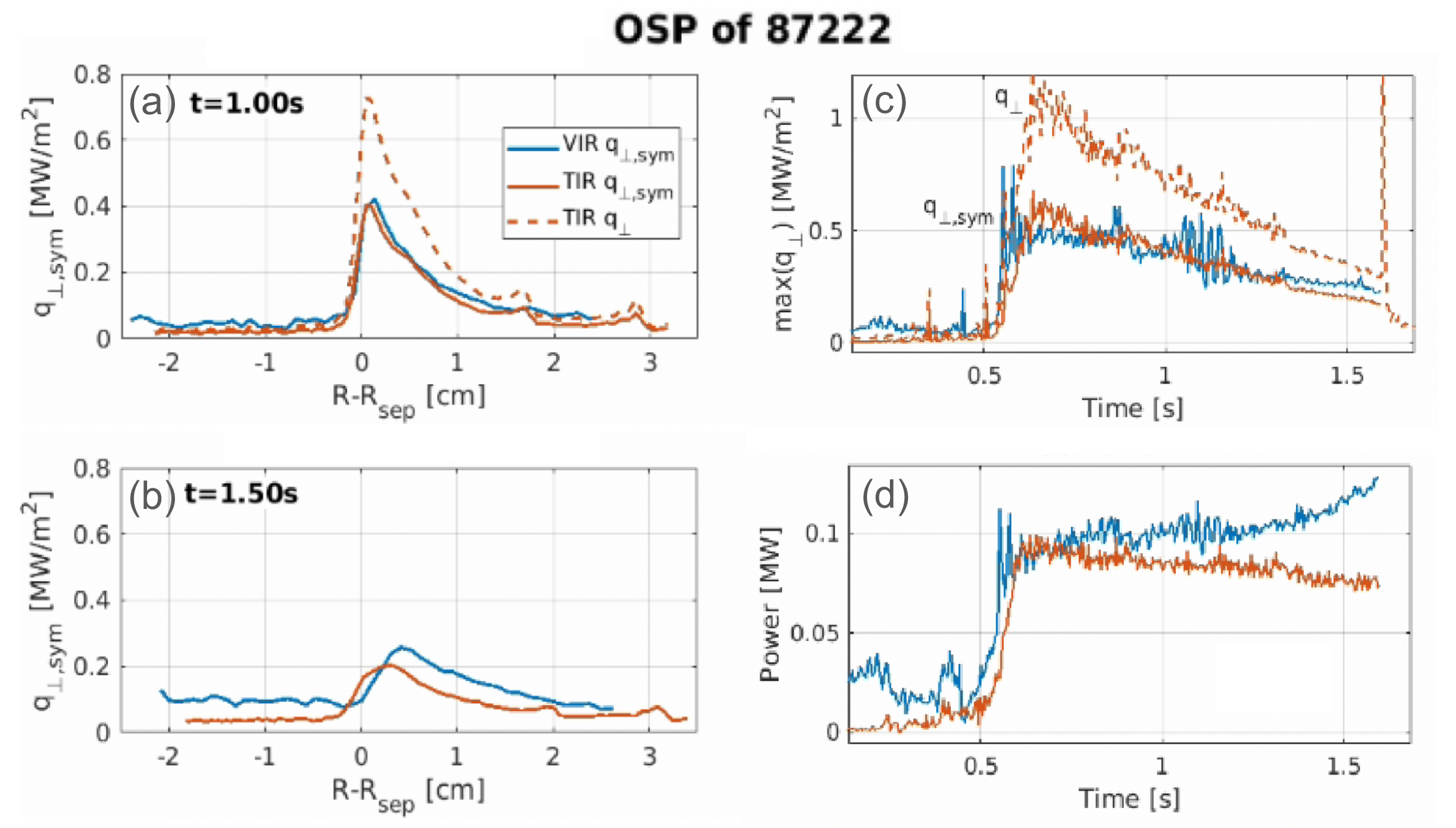}
\par\end{centering}
\caption{\label{fig: heat flux example}(a) Perpendicular heat flux profiles
at $t=1.00\;\text{s}$ for VIR and TIR in the discharge 87222 (same
as Fig. \ref{fig:HIR =000026 VIR view} and Fig. \ref{fig: TIR view}).
Solid lines: perpendicular heat flux projected in a toroidally symmetric
tile, using Eq. (\ref{eq: qperp sym def}). Dashed line: $q_{\perp}$
obtained with THEODOR, higher than $q_{\perp,sym}$ because of the
increased grazing angle of the TIR rooftop tile (Sec. \ref{subsec:TIR-rooftop-tiles}).
(b) $q_{\perp,sym}$ at $t=1.50\;\text{s}$. (c) Peak of the heat
flux profiles as a function of time. (d) Target power evaluated with
Eq. (\ref{eq: power def}).}
\end{figure}

Examples of perpendicular heat flux profiles measured by the VIR and
TIR systems are shown in Fig. \ref{fig: heat flux example} for the
outer strike point of discharge 87222 (IR thermogram in Fig. \ref{fig:HIR =000026 VIR view}(c)
and Fig. \ref{fig: TIR view}(b)). Due to the increased grazing angle
$\alpha$ of the TIR rooftop tile (Sec. \ref{subsec:TIR-rooftop-tiles}),
the TIR $q_{\perp}$ is considerably higher than its $q_{\perp,sym}$
(Fig. \ref{fig: heat flux example}(a) and (c)). Nevertheless, projecting
the heat flux for a toroidally symmetric tile using Eq. (\ref{eq: qperp sym def})
gives VIR and TIR heat fluxes that agree well at low plasma density.

The target peak heat flux (Fig. \ref{fig: heat flux example}(c))
decreases with time as the discharge 87222 had a density ramp, which
evolves the plasma towards a detached state. Nevertheless, it is seen
that the target power evaluated with Eq. (\ref{eq: power def}) disagrees
for VIR and TIR before the strike point arrives at the tile ($t<0.5\;\text{s}$)
and after $t>1.0\;\text{s}$, when the VIR power increases while the
TIR power decreases. This is caused by infrared plasma radiation measured
by the IR cameras, which affects especially the VIR (as discussed
in Sec. \ref{sec:IR-plasma-radiation} and in further detail in Sec.
4.5 of \cite{Colandrea 2024 PhD thesis}). As the density increases
(Thomson scattering separatrix density $n_{e,sep}=2\times10^{19}/\text{m}^{3}$
at $t=1.00\;\text{s}$ and $n_{e,sep}=4\times10^{19}/\text{m}^{3}$
at $t=1.50\;\text{s}$ for 87222), the target plasma temperature lowers,
increasing the plasma infrared emission, which is perceived by the
VIR camera as in increase in the tile temperature. This causes the
VIR estimated background heat flux and power to be overestimated (Fig.
\ref{fig: heat flux example}(b) and (d)).

\section{\label{sec:Summary}Summary}

TCV's infrared thermography system is composed by three IRCAM Equus
81k M cameras, HIR, VIR, and TIR, whose main views map the inner wall,
floor and tilted tiles of the tokamak. The camera raw data is calibrated
to temperature using heated tiles and thermocouples. The temperature
measurements are translated into target heat flux with the THEODOR
code. New VIR valley and TIR rooftop tiles have been commissioned
for fast transient studies.

The mass density, specific heat, diffusivity, and conductivity of
TCV's graphite tiles have been measured by NPL-UK in 2023 and 2025.
The mass density, identified between $1.823$ and $1.855\;\text{g}/\text{cm}^{3}$
can be considered constant, decreasing by less then 1\% from 20 ºC
to 500 ºC. The specific heat increased with temperature, going from
going from 0.7 J/g/K at 20 ºC to 1.6 J/g/K at 500 ºC. The diffusivity
and conductivity agree with the functional dependence expected by
THEODOR, yielding $D_{fit}(T)=\left\{ 10.6(2.7)\;+72.9(2.9)/\left[1+T/(392(56)\;\text{ºC})\right]^{2}\right\} \text{mm}^{2}/\text{s}$
and $\ensuremath{k_{fit}(T)=104.2(1.7)/\left[1+T/\left(2.82(0.19)\times10^{3}\;\text{ºC}\right)\right]\frac{\text{W}}{\text{m}\;\text{K}}}.$

The infrared plasma radiation measured by the Equus cameras has been
reduced with long wave pass filters, mainly to avoid the deuterium
line $5\rightarrow4$ at $4051\;\text{nm}$. However, at high plasma
density the infrared plasma emission still distorts the IR camera
signals in TCV, specially for the VIR system. This, together with
the determination of the surface layer heat transmission factor $\alpha_{top}$,
remain as the main sources of uncertainty for IR systems in TCV. Externally
heating a tile should decrease the effect of parasitic infrared radiation,
which is subject of future work.

\section{Acknowledgements }

The authors thank Michael Faitsch and Dirk Stieglitz for insightful
discussions and continued support with THEODOR. This work has been
carried out within the framework of the EUROfusion Consortium, partially
funded by the European Union via the Euratom Research and Training
Programme (Grant Agreement No 101052200 --- EUROfusion). The Swiss
contribution to this work has been funded in part by the Swiss State
Secretariat for Education, Research and Innovation (SERI). Views and
opinions expressed are however those of the author(s) only and do
not necessarily reflect those of the European Union, the European
Commission or SERI. Neither the European Union nor the European Commission
nor SERI can be held responsible for them.

\appendix

\section{\label{sec:Appendix:NPL tables}Appendix: Tables with the NPL tile
data}

Tables \ref{tab: NPL 2023} and \ref{tab: NPL 2025} contain the thermal
properties of TCV's graphite tiles measured by NPL-UK, used in Fig.
\ref{fig:NPL}.

\begin{table}

\caption{\label{tab: NPL 2023}Mass density, specific heat, diffusivity, and
conductivity for the graphite 2023 samples discussed in Sec. \ref{subsec: NPL}.}

\begin{centering}
\begin{tabular}{|c|c|c|c|c|}
\hline 
Temp {[}ºC{]} & $\rho\;[\text{g/cm}^{3}]$ & $c_{p}\;\text{[J/g/K]}$ & $D\;[\text{mm}^{2}/\text{s}]$ & $k\;[\text{J/m/K}]$\tabularnewline
\hline 
\hline 
18 & 1.836 & 0.696 & 82.97 & 106\tabularnewline
\hline 
99  & 1.834 & 0.916 & 61.14 & 103\tabularnewline
\hline 
299  & 1.829 & 1.350 & 35.31 & 87.2\tabularnewline
\hline 
500  & 1.823 & 1.616 & 25.67 & 75.6\tabularnewline
\hline 
400  & 1.827 & 1.500 & 29.07 & 79.7\tabularnewline
\hline 
202 & 1.832 & 1.141 & 43.14 & 90.2\tabularnewline
\hline 
17 & 1.836 & 0.696 & 81.00 & 104\tabularnewline
\hline 
\end{tabular}
\par\end{centering}
\textcompwordmark{}

\caption{\label{tab: NPL 2025}Mass density, specific heat, diffusivity, and
conductivity for the graphite 2025 samples.}

\begin{centering}
\begin{tabular}{|c|c|c|c|c|}
\hline 
Temp {[}ºC{]} & $\rho\;[\text{g/cm}^{3}]$ & $c_{p}\;\text{[J/g/K]}$ & $D\;[\text{mm}^{2}/\text{s}]$ & $k\;[\text{J/m/K}]$\tabularnewline
\hline 
\hline 
15 & 1.855 & 0.707 & 74.52 & 97.7\tabularnewline
\hline 
20 & 1.855 & 0.722 & 72.94 & 97.7\tabularnewline
\hline 
100 & 1.853 & 0.956 & 54.56 & 96.7\tabularnewline
\hline 
300 & 1.848 & 1.398 & 32.35 & 83.6\tabularnewline
\hline 
500 & 1.842 & 1.658 & 23.84 & 72.8\tabularnewline
\hline 
399 & 1.845 & 1.557 & 27.91 & 80.2\tabularnewline
\hline 
200 & 1.851 & 1.203 & 41.81 & 93.1\tabularnewline
\hline 
20 & 1.855 & 0.722 & 73.17 & 98.0\tabularnewline
\hline 
\end{tabular}
\par\end{centering}
\end{table}

\section{\label{sec:Appendix-C TIR rooftop}Appendix: Derivation of the TIR
rooftop tile surface}

\subsection{The rooftop surface equation in the tokamak frame}

\begin{figure}
\begin{centering}
\subfloat[]{\begin{centering}
\includegraphics[scale=0.4]{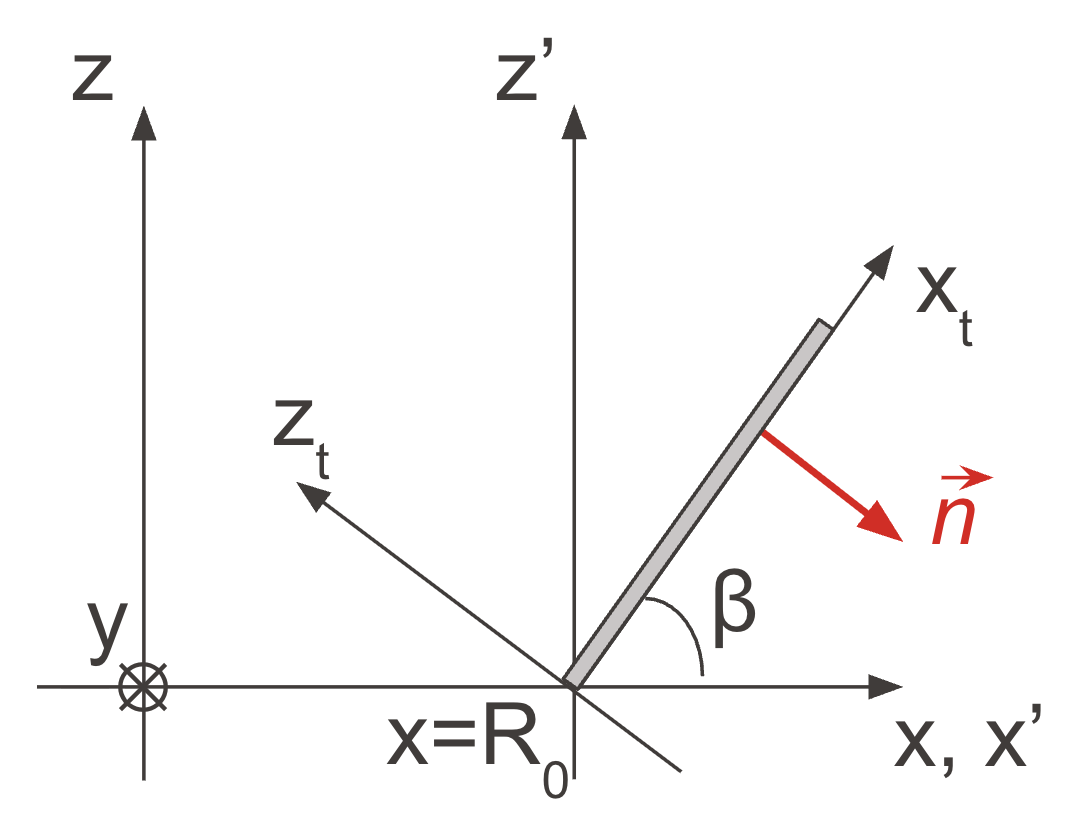}
\par\end{centering}
}\subfloat[]{\begin{centering}
\includegraphics[scale=0.5]{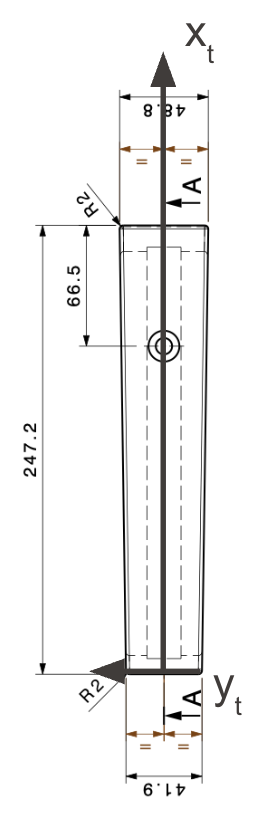}
\par\end{centering}
}
\par\end{centering}
\caption{\label{fig:TIR-rooftop-refs}(a) The three reference frames ($(x,y,z)$,
$(x',y',z')$, and $(x_{t},y_{t},z_{t})$) used to calculate the TIR
rooftop tile equation. Note that $z=z'$, $y=y'=y_{t}$, and that
$\beta=50\text{º}$ for the tilted tiles. (b) Dimensions of the tilted
tile {[}mm{]}.}
\end{figure}

The surface of the simple (non-rooftop) tilted tiles satisfies
\[
z(x,y)=\left(x-R_{0}\right)\tan\beta
\]
where $z$ is the height of the tile, $R_{0}=970\;\text{mm}$ is the
major radius at which the tilted tiles start and $\beta=50\text{º}$
is the tilted tiles angle with the floor. The reference frames used
in derivation and the dimensions of the tile are described in Fig.
\ref{fig:TIR-rooftop-refs}.

A toroidally symmetric tile would follow
\[
z=(R-R_{0})\tan\beta
\]
where $R$ is the major radius. Adding a component proportional to
$R\varphi$ forms a rooftop and increases the grazing angle,
\begin{equation}
z=(R-R_{0})\tan\beta+KR\varphi\label{eq: rooftop eq exact}
\end{equation}
where $\varphi$ is the toroidal angle (set to 0 at the center of
the tile) and $K$ is a constant set to increase a grazing angle of
$\alpha_{0}=4\text{º}$ (without the rooftop) to $\alpha=7\text{º}$.

\subsection{Finding the rooftop constant K}

To find $K$ in Eq. (\ref{eq: rooftop eq exact}), it is necessary
to evaluate the grazing angle (and hence the vector normal to the
surface). Let
\[
f(R,\varphi,z)=(R-R_{0})\tan\beta+KR\varphi-z=0
\]
Then vector normal to the surface is
\begin{align*}
\vec{n} & =\nabla f(R,\varphi,z)=\frac{\partial f}{\partial R}\hat{R}+\frac{\partial f}{R\partial\varphi}\hat{\varphi}+\frac{\partial f}{\partial z}\hat{z}\\
 & =(\tan\beta+K\varphi)\hat{R}+K\hat{\varphi}-\hat{z}
\end{align*}
and the unit normal vector is
\[
\hat{n}=\frac{\vec{n}}{\left|\vec{n}\right|}=\frac{(\tan\beta+K\varphi)\hat{R}+K\hat{\varphi}-\hat{z}}{\left[\left(\tan\beta+K\varphi\right)^{2}+K^{2}+1\right]^{1/2}}
\]
Assuming $K\varphi\ll\tan\beta$ and $K^{2}\ll1$,
\[
\hat{n}\approx\frac{(\tan\beta+K\varphi)\hat{R}+K\hat{\varphi}-\hat{z}}{\left(\tan^{2}\beta+1\right)^{1/2}}=\cos\beta\left(\tan\beta\hat{R}+K\hat{\varphi}-\hat{z}\right)
\]
The sine of the grazing angle between the plasma and the tile is
\[
\sin\alpha=\frac{\vec{B}}{B}\cdot\hat{n}
\]
where the magnetic field is
\[
\vec{B}=B_{R}\hat{R}+B_{\varphi}\hat{\varphi}+B_{z}\hat{z}
\]
So
\begin{align*}
\sin\alpha & =\frac{1}{B}\cos\beta\left(B_{R}\tan\beta+KB_{\varphi}-B_{z}\right)\\
 & =\sin\alpha_{0}+\frac{B_{\varphi}}{B}K\cos\beta
\end{align*}
where $\sin\alpha_{0}=\frac{1}{B}\cos\beta(B_{R}\tan\beta-B_{z})$
is the grazing angle sine for $K=0$, i.e. without the rooftop. Hence
\[
\text{\fbox{\ensuremath{K=\frac{B}{B_{\varphi}}\frac{\sin\alpha-\sin\alpha_{0}}{\cos\beta}}}}
\]
Assuming $B_{\varphi}=0.998B$, $\alpha=7\text{º}$, $\alpha=4\text{º}$
and $\beta=50\text{º}$ gives
\[
K=0.0812
\]
Note that since $B/B_{\varphi}\approx1$ and $\alpha\ll1$, a floor
tile (i.e. $\beta=0$) would have $K\approx\alpha-\alpha_{0}$.

\subsection{TIR rooftop tile surface in the tile frame}

The maximum toroidal angle variation of a tilted tile in TCV is $\varphi_{max}=1.24\text{º}\cdot\pi/180\text{º}=0.02164$.
Since $\varphi_{max}\ll1$, then $\cos\varphi_{max}=0.9998\approx1$
and $\sin\varphi_{max}=0.02164\ll1$. To second order in $\varphi$
and $y/x$,
\[
y=R\sin\varphi\approx R\varphi
\]
\[
R=\sqrt{x^{2}+y^{2}}=x\sqrt{1+y^{2}/x^{2}}\approx x\left(1+\frac{y^{2}}{2x^{2}}\right)=x+\frac{y^{2}}{2x}
\]
where $y_{max}/x_{0}=21/970=0.0216$ and thus $\left(y_{max}/x_{0}\right)^{2}=4.7\times10^{-4}$.
Therefore a second order approximation to the tile surface is
\[
z=\left(R-R_{0}\right)\tan\beta+KR\varphi\implies
\]
\[
\boxed{z\approx\left(x+\frac{y^{2}}{2x}-R_{0}\right)\tan\beta+Ky}
\]

It can be seen from a drawing that $z=z'=x_{t}\sin\beta+z_{t}\cos\beta$
and $z_{t}=-x'\sin\beta+z\cos\beta=-(x-R_{0})\sin\beta+z\cos\beta$
so
\[
z\cos\beta=z_{t}+(x-R_{0})\sin\beta
\]
and thus
\[
z\approx\left(x+\frac{y^{2}}{2x}-R_{0}\right)\tan\beta+Ky\implies
\]
\[
z\cos\beta=z_{t}+(x-R_{0})\sin\beta\approx\left(x+\frac{y^{2}}{2x}-R_{0}\right)\sin\beta+Ky\cos\beta\implies
\]
\[
z_{t}=\frac{y^{2}}{2x}\sin\beta+Ky\cos\beta
\]
Using $x-R_{0}=x'=x_{t}\cos\beta-z_{t}\sin\beta$,
\[
z_{t}=\frac{1}{2}\frac{y^{2}}{R_{0}+x_{t}\cos\beta-z_{t}\sin\beta}\sin\beta+Ky\cos\beta
\]
Since $z_{t}\sin\beta\ll R_{0}$,
\[
z_{t}\approx\frac{1}{2}\frac{y^{2}}{R_{0}+x_{t}\cos\beta}\sin\beta+Ky\cos\beta
\]
To manufacture the tile it is easier if $z_{t}$ only depends on one
coordinate. For this reason it is useful to approximate $y^{2}/(R_{0}+x_{t}\cos\beta)\approx y^{2}/R_{0}$.
This gives the expression used to manufacture the tile,
\[
\boxed{z_{t}(y)\approx\frac{y^{2}}{2R_{0}}\sin\beta+Ky\cos\beta}
\]
A slightly more precise approximation would be $y^{2}/(R_{0}+x_{t}\cos\beta)\approx y^{2}/(R_{0}+0.5x_{t,max}\cos\beta)$.

It is possible to check if $y^{2}/(R_{0}+x_{t}\cos\beta)\approx y^{2}/R_{0}$
is reasonable. Using $y(R_{0},\varphi_{max})=41.9\;\text{mm}$, $y(R_{max},\varphi_{max})=48.8\;\text{mm}$
(Fig. \ref{fig:TIR-rooftop-refs}(b)), $x_{t}(R_{max})=247.2\;\text{mm}$,
and $R_{0}\approx970\;\text{mm}$,
\[
\frac{1}{2}\frac{y^{2}(R_{0},\varphi_{max})}{R_{0}}=\frac{1}{2}\frac{41.9^{2}}{970}\;\text{mm}=0.90\;\text{mm}
\]
while
\[
\frac{1}{2}\frac{y^{2}(R_{0},\varphi_{max})}{R_{0}+x_{t}(R_{max})\cos\beta}=\frac{1}{2}\frac{48.8^{2}}{970+247.2\cos45\text{º}}\;\text{mm}=1.04\;\text{mm}
\]
So the maximum error is of $1.04\;\text{mm}-0.90\;\text{mm}=0.14\;\text{mm}$.
Furthermore, using, $y(R_{max},\varphi_{max})=48.8\;\text{mm}$, $x_{t}(R_{max})=247.2\;\text{mm}$,
and $R_{0}\approx970\;\text{mm}$ gives
\[
\frac{1}{2}\frac{y^{2}(R_{max})}{R_{0}+x_{t}(R_{max})\cos\beta}\sin\beta=\frac{1}{2}\frac{48.8^{2}\sin45\text{º}}{970+247.2\cos45\text{º}}\;\text{mm}=0.74\;\text{mm}
\]
while the approximation is
\[
\frac{1}{2}\frac{y^{2}(R_{max})}{R_{0}}\sin\beta=\frac{1}{2}\frac{48.8^{2}\sin45\text{º}}{970}\;\text{mm}=0.87\;\text{mm}
\]
Therefore the highest error committed is of $0.87\;\text{mm}-0.74\;\text{mm}=0.13\;\text{mm}$
which is close to the machining precision ($0.1\;\text{mm}$).

\begingroup 
\global\long\def\thefootnote{\fnsymbol{footnote}}%

\section*{The TCV team\protect\footnote{As in B. P. Duval et al. Nucl. Fusion 64 112023 (2024).}}

\endgroup A. Abdolmaleki$^{2}$, M. Agostini$^{3}$, C.J. Ajay$^{4}$,
S. Alberti$^{1}$, E. Alessi$^{5}$, G. Anastasiou$^{6}$, Y. Andrèbe$^{1}$,
G.M. Apruzzese$^{7}$, F. Auriemma$^{3}$, J. Ayllon-Guerola$^{8}$
% ...
\\

\noindent $^{1}$ École Polytechnique Fédérale de Lausanne (EPFL),
Swiss Plasma Center (SPC), Lausanne, Switzerland\\
 $^{2}$ Google DeepMind, London, United Kingdom\\
 $^{3}$ Consorzio RFX, Padova, Italy\\
 $^{4}$ York Plasma Institute, University of York, Heslington, York,
United Kingdom\\
 $^{5}$ Istituto per la Scienza e Tecnologia dei Plasmi ISTP-CNR,
Milano, Italy\\
 $^{6}$ Aristotle University of Thessaloniki, Thessaloniki, Greece\\
 $^{7}$ Unità Tecnica Fusione, ENEA, Frascati, Italy\\
 $^{8}$ Universidad de Sevilla, Sevilla, Spain\\
% ...
\end{document}